\documentstyle[12pt,epsf]{article}

\newcommand{\e}{\mbox{\boldmath $\epsilon$}}
\newcommand{\te}{\mbox{\boldmath $\widetilde{\epsilon}$}}
\newcommand{\I}{{\rm I}}
\newcommand{\II}{{\rm II}}
\newcommand{\tI}{\widetilde{\rm I}}
\newcommand{\tII}{\widetilde{\rm II}}

\newcommand{\bx}{\bar{x}}
\newcommand{\by}{\bar{y}}

\newcommand{\be}{\bar{\eta}}
\newcommand{\bi}{\bar{\xi}}
\newcommand{\teta}{\tilde{\eta}}
\newcommand{\we}{{w_{\perp}}}
\newcommand{\lp}{{\lambda_{\perp}}}
\newcommand{\X}{\bar{X}}
\newcommand{\ch}{\widehat{c}}
\newcommand{\sh}{\widehat{s}}
\newcommand{\p}{|{\bf P}|}

\begin{document}

\title{Bound and Radiation Fields in the Rindler Frame}

\author{Toru {\sc Hirayama}\thanks{E-mail: hira@cc.kyoto-su.ac.jp} 
\\
\small Department of Physics, Kyoto Sangyo University, 
Kyoto 603-8555, Japan}
\date{August, 2001}

\maketitle 

\begin{abstract}

The energy-momentum tensor of the Li\'enard-Wiechert field is split into 
bound and emitted parts in the Rindler frame, by generalizing the reasoning 
of Teitelboim applied in the inertial frame 
$[$see C. Teitelboim, Phys. Rev. {\bf D1} (1970), 1572$]$.  Our analysis 
proceeds by invoking the concept of ``energy'' defined with respect 
to the Killing vector field attached to the frame.
We obtain the radiation formula in the Rindler frame (the Rindler version of 
the Larmor formula), and it is found that the radiation power is proportional 
to the square of acceleration $\alpha^\mu$ of the charge relative to 
the Rindler frame.
This result leads us to split the Li\'enard-Wiechert field into a 
part $\tII$, which is linear in $\alpha^\mu$, and a part $\tI$, which 
is independent of $\alpha^\mu$. By using these, we split the 
energy-momentum tensor into two parts. We find that these  
are properly 
interpreted as the emitted and bound parts of the tensor 
in the Rindler frame.
In our identification of radiation, a charge radiates neither 
in the case that the charge is fixed in the Rindler frame, nor 
in the case that the charge satisfies the equation $\alpha^\mu=0$. 
We then investigate this equation. We consider four 
{\it gedanken experiments}  
related to the observer dependence of the 
concept of radiation.  

\end{abstract}

\section{Introduction}
\label{Introduction}

Li\'enard-Wiechert fields, which represent the electromagnetic field 
generated by a point charge,  consist of 
a part which is linear in the acceleration $a^\mu$ of a point charge, 
and a part which is independent of $a^\mu$. The latter part, 
which is called ``velocity fields'', decreases as $R^{-2}$ with increasing 
distance $R$ from the charge, and is considered to be bound to 
the charge (a generalized Coulomb field). The former part, which 
is called ``acceleration fields'', decreases as $R^{-1}$, and is 
considered to be emitted from the charge.

Rohrlich~\cite{Ro61} showed that one can define radiation at any 
distance from a charge without referring to the region of 
large $R$ (wave zone), and this point of view was further developed 
by Teitelboim.
Teitelboim~\cite{Te70,TV80} split the energy-momentum tensor of 
retarded fields into a part $\II$, which is related only to 
acceleration fields, and a part $\I$, which consists of the 
contribution of the pure velocity fields and the contribuiton of 
the interference between velocity fields and acceleration fields. 
He found that part $\II$ propagates along the future light 
cone while its energy and momentum do not decrease. 
Therefore, this part is regarded as the emitted part of the tensor 
at any distance from the charge, and one need not investigate 
the behavior of the field in the wave zone to identify the radiation.

The definition of radiation given by Rohrlich and Teitelboim is 
appropriate  in inertial frames (see Section 5 of Ref.~\cite{Ro61}), 
and their definition concerns the detection of radiation 
by an observer who experiences inertial motion. However, one can 
consider the situation that an observer detects radiation while 
he experiences accelerated motion, and such consideration might 
help us investigate the definition of radiation in general 
curved spacetimes. For this reason, in this paper, we attempt to identify the 
emitted part and bound part of the elecromagnetic field generated 
by a point charge in the Rindler frame,\footnote{An alternative 
approach to generalize the work of Rohrlich and Teitelboim to 
accelerated frames was previously performed in 1991 by Fugmann 
and Kretzschmar~\cite{FK91}.  Let us compare our work here with 
their work.
Our work proceeds by using the time-like Killing vector 
field and the radiation formula in the Rindler frame. Contrastingly, 
they attached the advanced optical coordinates to the 
observer with arbitrary acceleration and rotation ( not 
restricting the accelerated frames to the frames generated by Killing 
vector fields), and identified the emitted part by extracting the 
contribution that is proportional to $r^{-2}$ in the elctromagnetic 
flux generated by the charge, where $r$ is the radial coordinate of 
the advanced optical coordinates, which represents the distance 
between the observer and the charge. Comparing our result with their 
result (by applying Eq.~(6.9) in Ref.~\cite{FK91} to the Rindler frame), 
we find that our result is consistent with their result when the 
charge is instantaneously at rest in the Rindler frame (that is, in 
the case $v^\mu=u^\mu$),  but in more general situations, our result 
is not consistent with their result. Details of the comparison of the two 
works are given in Appendix \ref{comparison}. } which is interpreted as a 
uniformly accelerated frame in Minkowski spacetime, and is regarded 
as an example of a curved spacetime (a constant homogeneous 
gravitational field).

We begin our discussion in a more general context than that restricted 
to the Rindler frame. In Minkowski spacetime, we consider the definition 
of radiation in an accelerated frame with a time-like Killing vector 
field. (we call this frame 
``the stationary frame'' \footnote{This is an analogue of the term 
``stationary spacetime'' (see Section 6.1 in Ref.~\cite{Wa84}).} 
for simplicity.)
In a stationary frame, one can define energy by invoking the invariance 
with respect to time translation generated by Killing vector fields.
By using this concept of energy, we generalize the reasoning of 
Teitelboim to stationary frames. To find the proper identification 
of the emitted part and bound part, we introduce the following three 
conditions as guiding principles. (1) A charge fixed in the frame 
dose not radiate. (2) The energy of the emitted part propagates along 
the future light cone without damping. (3) The bound part does not 
contribute to the energy at the infinity from the charge (see 
\S \ref{Kgeneral}).

Next, we apply the result obtained in stationary frames to the Rindler 
frame. By using the third condition mentioned above, one finds 
that only the emitted part contributes to the Rindler energy in the 
region infinitely distant from the charge.
Then we evaluate the energy (of course, this is defined in 
the Rindler frame) of the total electromagnetic field in this region, 
and obtain the radiation power emitted per unit proper time of the 
charge  $[$Eq.~(\ref{formula})$]$. We find that the result is 
proportional to the square of the acceleration 
$\alpha^\mu=h^\mu_{\ \nu}[a^\nu-(g^\alpha g_\alpha)^{1/2}u^\nu-g^\nu]$
relative to the Rindler frame, where 
$h^\mu_{\ \nu}=\delta^\mu_{\ \nu}+v^\mu v_\nu$
is a projector, $v^\mu$ and $a^\mu$ are the 4-velocity 
and 4-acceleration of the charge, and $u^\mu$ and $g^\mu$ are  
the 4-velocity and 4-acceleration of a Rindler observer. This 
resembles the situation in inertial frames, where the Larmor 
formula is proportional to the square of the acceleration $a^\mu$. 
This result leads us to split the retarded fields into a part 
$\tII$ that is linear in $\alpha^\mu$ and a part $\tI$ 
that is independent of $\alpha^\mu$, and this causes the 
splitting of the energy-momentum tensor of the field into the 
emitted part $\tII$ and bound part $\tI$. We find that this 
splitting of the tensor satisfies our three conditions.

In our result, a charge radiates neither in the case 
that the charge is fixed to the Rindler frame, nor in the 
case that the trajectory of the charge satisfies the equation 
$\alpha^\mu=0$. Therefore we solve this equation and investigate the 
properties of this trajectory. Also we briefly investigate the 
action of the charge on the bound part $\tI$ in terms of 
the Rindler energy.

There is an old paradox of classical electrodynamics 
concerning whether a uniformly accelerated charge 
radiates~\cite{FR60,Bo80}.  We briefly interpret our result 
by applying it to four {\it gedanken experiments} related 
to this paradox.    

After this section, we proceed as follows.  
In \S~\ref{splitting}, we generalize Teitelboim's work to 
stationary frames, and by using this and the radiation formula 
in the Rindler frame derived in Appendices \ref{appendix} and \ref{new}, 
we identify the emitted 
part and the bound part in the Rindler frame. In \S~\ref{motion}, 
we solve the equation $\alpha^\mu=0$ in the Rindler frame, and 
investigate its physical and geometrical properties. We also 
make a brief calculation concerning the energy of the bound part. 
In \S~\ref{discussion}, we discuss our result and briefly 
comment on the radiation paradox. In Appendix \ref{comparison}, our 
work is compared with the work of Fugmann and Kretzschmar. Throughout 
the paper, we use 
Gaussian units and the metric with signature $(-,+,+,+)$.

\section{Identification of bound and radiation fields}
\label{splitting}

Although the purpose of this paper is to determine the proper 
identification of the bound and emitted parts of the Maxwell tensor 
in the Rindler frame, it would be useful to proceed in as general 
a context as possible for the purpose of future generalization 
of the work. For this reason, we begin our discussion of the general case of 
accelerated frames that are generated by time-like Killing vector 
fields. Later, we restrict our attention to the special case of 
the Rindler frame.

\subsection{Energy generated by the Killing vector field}
\label{secKilling}

Because we seek the proper identification of emitted parts 
in accelerated frames, it seems natural to evaluate the radiation 
power with respect to the energy defined in accelerated frames, 
not in inertial frames. We restrict our 
consideration to accelerated frames that are generated by time-like 
Killing vector fields (stationary frames), because in such frames 
one can introduce the concept of energy by invoking invariance 
with respect to time translations generated by the Killing fields.
(The concept of energy introduced here is used in 
recent papers~\cite{Pa93} and~\cite{SK99}, concerning classical 
radiation in the Rindler frame.)

Let $X^\mu$ be a Killing field attached to a stationary frame.
For an energy-momentum tensor that satisfies 
$\nabla_\mu T^{\mu\nu}=0$, the conservation law
\begin{eqnarray}
\int_{\dot{N}}\bar{n}_\mu X_\nu T^{\mu\nu} \te
&=&0
\label{hozon}
\end{eqnarray}
follows. $[\dot{N}$ is the 3-dimensional boundary of the 
4-dimensional region $N$. If $\dot{N}$ is not a null surface, 
$\bar{n}^\mu$ is a unit normal to $\dot{N}$, and $\te$ is a volume 
element of $\dot{N}$. In this case,  $\bar{n}_\mu$ is directed to 
the outside of the region $N$. If $\dot{N}$ is a null surface, 
$\bar{n}^\mu$ is an arbitrary vector which is orthogonal to 
surface $\dot{N}$, and $\te$ is a 3-form such that for a 
4-dimensional volume element $\e$, it satisfies 
$\e=\bar{\bf n}\wedge\te$ ($\bar{\bf n}=\bar{n}_\mu dx^\mu$).$]$
This can be proved using Gauss's theorem
(see Appendix B in Ref.~\cite{Wa84}),
\begin{eqnarray}
\int_{{\rm int}(N)}\nabla_\mu f^\mu \e
&=&
\int_{\dot{N}}\bar{n}_\mu f^\mu \te,
\label{gauss}
\end{eqnarray}
where ${\rm int}(N)=N-\dot{N}$, in the following way. 
We set $f^\mu=X_\nu T^{\mu\nu}$ in Eq.~(\ref{gauss}). 
Then the integrand on the left-hand side of Eq.~(\ref{gauss}) gives
\begin{eqnarray}
\nabla_\mu f^\mu &=&
(\nabla_\mu X_\nu) T^{\mu\nu} + X_\nu (\nabla_\mu T^{\mu\nu})\nonumber\\
&=&(\nabla_\mu X_\nu) T^{\mu\nu}\nonumber\\
&=&-(\nabla_\nu X_\mu) T^{\mu\nu}\nonumber\\
&=&-(\nabla_\nu X_\mu) T^{\nu\mu}\nonumber\\
&=& 0,
\end{eqnarray}
where we have used the Killing equation 
$\nabla_\mu X_\nu +\nabla_\nu X_\mu = 0$, $\nabla_\mu T^{\mu\nu}=0$ 
and $T^{\mu\nu}=T^{\nu\mu}$, so that we obtain Eq.~(\ref{hozon}).

Now we consider the quantity
\begin{eqnarray}
E_{{\rm K}[\Sigma]}&=&
-\int_\Sigma 
n_\mu X_\nu T^{\mu\nu}
\te,
\end{eqnarray}
where $\Sigma$ is a 3-dimensional space-like surface, $n^\mu$ 
is a unit normal to $\Sigma$, $T^{\mu\nu}$ is the energy-momentum 
tensor of the electromagnetic field, $\te=\sqrt{g_{_\Sigma}}d^3x$ 
is a volume element of $\Sigma$, and $g_{_\Sigma}$ is the 
determinant of the metric in the 3-dimensional region $\Sigma$. 
From Eq.~(\ref{hozon}), we find that, if the contribution of 
the electromagnetic field is sufficiently small in the infinite region, 
this quantity does not depend on the choice of $\Sigma$. That is, 
for any two distinct simultaneous surfaces $\Sigma_1$ and $\Sigma_2$, 
it follows that
\begin{eqnarray}
E_{{\rm K}[\Sigma_1]}&=&E_{{\rm K}[\Sigma_2]}.
\end{eqnarray}
Therefore, in this case, $E_{{\rm K}[\Sigma]}$ can be regarded as 
a conserved quantity. Since the conservation property is 
attributed to the invariance of the frame with respect to time 
translation, we can interpret this quantity as energy in the 
stationary frame.

\subsection{Radiation fields in the accelerated frame induced 
by the Killing vector field}
\label{Kgeneral}

\begin{figure}
\epsfxsize=6.2cm
\centerline{\epsfbox{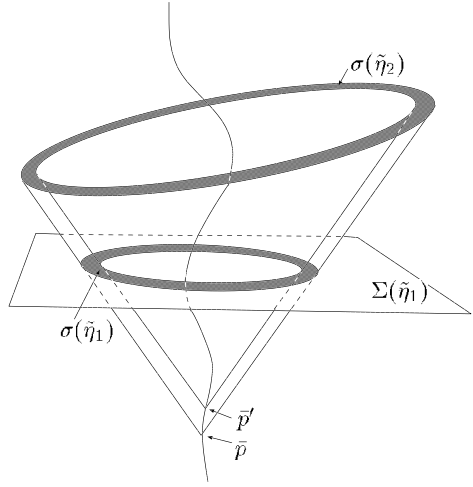}}
\caption{The emitted part propagates along the future light 
cone without losing energy.}
\label{nullnull}
\end{figure}

As in Fig. \ref{nullnull}, we take two closely neighboring 
points $\bar{p}$ and $\bar{p}'$ on the worldline of the charge, 
and consider the behavior of the electromagnetic field in the region  
$\Delta L$, which is bounded on the future light cones $L$ and $L'$ 
with apexes $\bar{p}$ and $\bar{p}'$.
Let $\Sigma(\tilde{\eta})$ be the space-like surface that consists 
of simultaneous points with respect to time defined in the 
accelerated frame with value $\teta$, and let $\sigma(\tilde{\eta})$ 
be the intersection of  $\Delta L$ with $\Sigma(\tilde{\eta})$.

As a guide to identify the bound and emitted parts properly in a 
stationary frame, we introduce three conditions which we assume 
to be satisfied in the proper identification:\\

\begin{enumerate}

\item[1.]{\it A charge fixed in the accelerated frame does not radiate}.\\

\item[2.]{\it The emitted part propagates along the future light cone 
with an apex at a fixed point on the world line of the charge, 
in the sense that the energy of this part does not damp 
along the light cone}. \\

\item[3.]{\it The bound part does not contribute to the energy in the 
region 
$\tilde{\eta}\rightarrow\infty$, where $\tilde{\eta}$ is the time 
coordinate in the accelerated frame and the limit is taken along 
the future light cone with an apex at a fixed point on the world line 
of the charge}.\\

\end{enumerate}

Condition 1 seems to be appropriate  in the Rindler frame, because, 
as is well known, the Poynting vector of the 
electromagnetic field  generated by a charge fixed in the Rindler 
frame is zero when evaluated by an observer fixed in 
the same frame~\cite{Bo80}.   
The validity of condition 1 in general stationary frames would 
also depend on the evaluation of the Poyinting vector generated 
by the charge fixed in the stationary frames.

The energy of the emitted part $E_{{\rm K}[\sigma]{\rm rad}}$ in the 
region $\sigma$ is given as 
\begin{eqnarray}
E_{{\rm K}[\sigma]{\rm rad}}&=&
-\int_{\sigma} 
n_\mu X_\nu {T_{\rm rad}}^{\mu\nu}
\te,
\end{eqnarray}
where ${T_{\rm rad}}^{\mu\nu}$ is the energy-momentum tensor of 
the radiation fields.
Condition 2 is expressed as
\begin{eqnarray}
E_{{\rm K}[\sigma(\tilde{\eta}_1)]{\rm rad}}
&=&E_{{\rm K}[\sigma(\tilde{\eta}_2)]{\rm rad}},
\label{yurui}
\end{eqnarray}
where $\tilde{\eta}_1\ne \tilde{\eta}_2$ (see Fig. \ref{nullnull}).
As mentioned in Teitelboim's paper~\cite{Te70},  the part which 
satisfies condition 2 is interpreted as a wave emitted by the 
charge from the retarded point $\bar{p}$, propagating at the 
speed of light.

By assuming $\nabla_\mu {T_{\rm rad}}^{\mu\nu}=0$, we obtain, 
by using Eq.~(\ref{hozon}),
\begin{eqnarray}
& &E_{{\rm K}[\sigma(\tilde{\eta}_1)]{\rm rad}}-
E_{{\rm K}[\sigma(\tilde{\eta}_2)]{\rm rad}}\nonumber\\
& &\hspace{2.2cm}=
\int_{L[\tilde{\eta_1},\tilde{\eta_2}]}
\tilde{n}_\mu X_\nu {T_{\rm rad}}^{\mu\nu}\te+
\int_{L'[\tilde{\eta_1},\tilde{\eta_2}]}
\tilde{n}_\mu X_\nu {T_{\rm rad}}^{\mu\nu}\te,
\label{closed}
\end{eqnarray}
where $L[\tilde{\eta_1},\tilde{\eta_2}]$ and 
$L'[\tilde{\eta_1},\tilde{\eta_2}]$ are the portions of 
$L$ and $L'$ that are bounded by the surfaces 
$\teta=\tilde{\eta_1}$ and $\teta=\tilde{\eta_2}$ 
($\tilde{\eta_1}<\tilde{\eta_2}$).
Then we may adopt the mathematical expression of 
condition 2 
\begin{eqnarray}
\nabla_\mu {T_{\rm rad}}^{\mu\nu}&=&0\ \ \ {\rm and}\ \ \ 
\hat{n}_\mu X_\nu {T_{\rm rad}}^{\mu\nu}=0
\label{strong}
\end{eqnarray}
($\hat{n}_\mu$ is a null vector which is orthogonal to $L$ or $L'$),
which is a more restricted expression than Eq.~(\ref{yurui}).

The bound part of the energy-momentum tensor
\begin{eqnarray}
{T_{\rm bound}}^{\mu\nu}
&=&{T_{\rm ret}}^{\mu\nu}-{T_{\rm rad}}^{\mu\nu},
\end{eqnarray}
where ${T_{\rm ret}}^{\mu\nu}$ is the total energy-momentum 
tensor of retarded fields, consists of the contribution of 
the pure bound fields and the contribution of the interference 
between bound fields and radiation fields $[$see the definition 
of ${T_{\I}}^{\mu\nu}$ in Ref.~\cite{Te70} or Eq.~(\ref{deftII}) 
in the present paper$]$.
The energy of the bound part $E_{{\rm K}[\sigma]{\rm bound}}$ 
in the region 
$\sigma$ can also be expressed as 
\begin{eqnarray}
E_{{\rm K}[\sigma]{\rm bound}}&=&
-\int_\sigma 
n_\mu X_\nu {T_{\rm bound}}^{\mu\nu}
\te,
\end{eqnarray}
then condition 3 is expressed as
\begin{eqnarray}
\lim_{\tilde{\eta}\rightarrow\infty}
E_{{\rm K}[\sigma(\tilde{\eta})]{\rm bound}}&=&0.
\label{condbound}
\end{eqnarray}
The part which satisfies condition 3 can be interpreted as not 
escaping to the region infinitely separated from the charge, 
so we may regard this part as that ``bound to the charge''. Now, we 
write the total energy of the retarded fields as 
$E_{{\rm K}[\sigma]{\rm ret}}$. Then, with the relation 
$E_{{\rm K}[\sigma]{\rm rad}}=
E_{{\rm K}[\sigma]{\rm ret}}-E_{{\rm K}[\sigma]{\rm bound}}$
and with Eqs. (\ref{yurui}) and (\ref{condbound}), we find 
\begin{eqnarray}
E_{{\rm K}[\sigma]{\rm rad}}&=&
\lim_{\tilde{\eta}\rightarrow\infty}
E_{{\rm K}[\sigma(\tilde{\eta})]{\rm rad}}
=
\lim_{\tilde{\eta}\rightarrow\infty}
E_{{\rm K}[\sigma(\tilde{\eta})]{\rm ret}}.
\label{tokkakari}
\end{eqnarray}

In the following, we seek radiation 
fields that satisfy conditions 1 and 2 by generalizing 
the reasoning of Teitelboim for the acceleration fields  
$[$presented from Eq.~(2$\cdot$1) to Eq.~(2$\cdot$11) in Ref.~\cite{Te70}$]$.
We here note that condition 1 might not be valid in the case 
of general stationary frames. However, the consideration given 
below should be useful 
to examine the identification of radiation in the Rindler frame in 
a more general context.

For a point $p$ (with coordinate value $x^\mu$),
a retarded point $\bar{p}$ (with coordinate value $\bx^\mu$) is 
defined as the point on the worldline of the charge that satisfies 
the condition
\begin{eqnarray}
R^\mu&=&x^\mu-\bar{x}^\mu,
\label{inertia}
\\
R^\mu R_\mu&=&0\\
x^t&>&\bar{x}^t\nonumber
\end{eqnarray}
expressed in an inertial frame $(t,x,y,z)$.
Here we note that the definition of $R^\mu$ in Eq.~(\ref{inertia}) is 
invariant under Lorentz transformations, but it is not invariant under 
general coordinate transformations. 
Therefore,  we express the equations only in inertial frames if the equations 
include the notation $R^\mu$ explicitly, unless 
otherwise stated.

For an arbitrary vector $\lambda^\mu$ at the retarded point 
$\bar{p}$, we define a field
\begin{eqnarray}
f^{\mu\nu}(x)&=&
\frac{e}{R^3}(\lambda^\alpha R_\alpha)[v^\mu R^\nu-R^\mu v^\nu]
+\frac{e}{R^2}[\lambda^\mu R^\nu-R^\mu \lambda^\nu],
\label{lambda}
\end{eqnarray}
where $v^\mu$ is evaluated at the retarded point $\bar{p}$, and 
we set $R=-v^\mu R_\mu$. Here, $f^{\mu\nu}$ is invariant with respect 
to the transformation $\lambda^\mu\rightarrow\lambda^\mu+lv^\mu$ 
(where $l$ is an arbitrary factor). Hence, by using 
$\lp^\mu\equiv h^\mu_{\ \nu}\lambda^\nu$ 
(where $h^\mu_{\ \nu}=\delta^\mu_{\ \nu}+v^\mu v_\nu$ is the projector 
 on the plane orthogonal to $v^\mu$), we can rewrite 
the field as
\begin{eqnarray}
f^{\mu\nu}(x)&=&
\frac{e}{R^3}(\lp^\alpha R_\alpha)[v^\mu R^\nu-R^\mu v^\nu]
+\frac{e}{R^2}[\lp^\mu R^\nu-R^\mu \lp^\nu].
\label{lp}
\end{eqnarray}

Taking the partial derivative of $\bx^\mu$ with respect to $x^\nu$ gives
\begin{eqnarray}
\partial_{\nu}\bx^\mu&=&
v^\mu\partial_{\nu}\tau,
\label{hen}
\end{eqnarray}
where $\tau$ is the proper time of the charge. From Eq.~(\ref{hen}), 
we obtain
\begin{eqnarray}
0&=&\partial_\mu(R^\alpha R_\alpha)
=2[R_\mu+R(\partial_\mu\tau)],
\end{eqnarray}
so that
\begin{eqnarray}
\partial_\mu\tau&=&-\frac{R_\mu}{R}.
\label{tau}
\end{eqnarray}
Using this equation, we find
\begin{eqnarray}
\partial_\mu R^\nu&=&
\delta^\nu_{\ \mu}-\partial_\mu \bx^\nu
=\delta^\nu_{\ \mu}-v^\nu\partial_\mu\tau
=\delta^\nu_{\ \mu}+\frac{v^\nu R_\mu}{R},
\label{one}\\
\partial_\mu R&=&
\frac{1+a^\alpha R_\alpha}{R}R_\mu-v_\mu,
\label{two}\\
\partial_\mu \lambda^\nu&=&
(\partial_\mu\tau)\partial_\tau\lambda^\nu
=
-\frac{R_\mu(\partial_\tau\lambda^\nu)}{R}.
\label{three}
\end{eqnarray}
Using Eqs.  (\ref{one}), (\ref{two}) and (\ref{three}), 
we obtain the equations of the field $f^{\mu\nu}$ off the worldline 
of the charge,
\begin{eqnarray}
\partial_{[\lambda}f_{\mu\nu]}&=&0,
\label{kasekibun}\\
\partial_\mu f^{\mu\nu}&=&
-\frac{2e}{R^4}(\lp^\alpha R_\alpha)R^\nu.
\label{fieldeq}
\end{eqnarray}

The energy-momentum tensor of the field $f^{\mu\nu}$ is given 
in the form 
\begin{eqnarray}
t^{\mu\nu}&=&\frac{1}{4\pi}\left[
f^{\mu\alpha}f_\alpha^{\ \nu}+
\frac{1}{4}g^{\mu\nu}f_{\alpha\beta}f^{\alpha\beta}
\right].
\label{stress}
\end{eqnarray}
With Eq.~(\ref{kasekibun}) and $f_{\mu\nu}=-f_{\nu\mu}$, we find
\begin{eqnarray}
\partial_\mu t^{\mu\nu}&=&
\frac{1}{4\pi}f^{\alpha\nu}(\partial_\mu f^\mu_{\ \alpha}).
\end{eqnarray}
Substituting Eqs. (\ref{lambda}) and (\ref{fieldeq}) into the above 
equation, we obtain the conservation law 
$\partial_\mu t^{\mu\nu}=0$ off the worldline. This result can be  
rewritten in general curvilinear coordinates 
with the covariant derivative:
\begin{eqnarray}
\nabla_\mu t^{\mu\nu}&=&0.
\label{cons}
\end{eqnarray}
Substituting Eq.~(\ref{lp}) into (\ref{stress}) and using 
$\lp^\alpha v_\alpha=0$, we obtain
\begin{eqnarray}
t^{\mu\nu}&=&
\frac{e^2}{4\pi R^6}
[(\lp^\alpha R_\alpha)^2-R^2(\lp^\alpha \lp_\alpha)]R^\mu R^\nu.
\end{eqnarray}

From Eq.~(\ref{cons}) and by noting $\hat{n}^\mu\propto R^\mu$ 
and $t^{\mu\nu}\propto R^\mu R^\nu$, we find that 
$t^{\mu\nu}$ satisfies  Eq.~(\ref{strong}). 
Therefore the field $f^{\mu\nu}$ defined in the stationary frame 
generated by an arbitrary Killing field satisfies condition 2. 
If we choose $\lambda^\mu$ so that $\lp^\mu=0$ for the motion of the 
charge fixed in the frame, the field satisfies condition 1 also. 
For example, with the value $u^\mu$ at the point $\bar{p}$ of the 
4-velocity $\tilde{u}^\mu=(-X^\alpha X_\alpha)^{-1/2}X^\mu$ of an 
observer fixed in the frame, and with the value $g^\mu$ at the point 
$\bar{p}$ of the 4-acceleration 
$\tilde{g}^\mu=\tilde{u}^\alpha\nabla_\alpha \tilde{u}^\mu$, 
we can choose $\lambda^\mu$ in the form
\begin{eqnarray}
\lambda^\mu&=&a^\mu-g^\mu-ku^\mu,
\label{example}
\end{eqnarray}
where $k$ is an arbitrary scalar at the point $\bar{p}$.
In this case, we find that $\lp^\mu=0$ for a charge fixed in the 
frame ($v^\mu=u^\mu$, $a^\mu=g^\mu$), and therefore the field 
$f^{\mu\nu}$ with $\lambda^\mu$ defined as Eq.~(\ref{example}) 
satisfies conditions 1 and 2 which are required for the proper 
definition of the radiation fields.
Thus we find that, for any stationary frame, there are several possible ways 
to define radiation fields that satisfy conditions 1 and 2 
(at least, one can choose the form of $k$ freely).

\subsection{Bound and radiation fields in the Rindler frame}
\label{Rindlersec}

In this subsection, we identify the radiation fields and bound fields 
in the Rindler frame by using the results of previous subsection, and 
with this identification, we split the energy-momentum tensor into 
the emitted part and bound part.

For an inertial frame $(t,x,y,z)$, the Rindler frame $(\eta,x,y,\xi)$ 
(see Section 4.5 in Ref.~\cite{BD82}) is defined as
\begin{eqnarray}
t&=&\xi\sinh\eta,\nonumber\\
z&=&\xi\cosh\eta,
\end{eqnarray}
where $-\infty<\eta<\infty$, $0<\xi<\infty$. (The Rindler frame can also 
be defined in the region $-\infty<\xi<0$, but we consider $\xi>0$ for 
simplicity.) In this case, the Rindler frame covers the region with 
$|t|<z$ in Minkowski spacetime. The metric in the Rindler frame 
is given as
\begin{eqnarray}
ds^2&=&
-\xi^2d\eta^2+dx^2+dy^2+d\xi^2.
\end{eqnarray}
Since the metric is independent of time $\eta$, there are Killing 
fields $X=\nu\partial_\eta$ ($\nu$ is an arbitrary constant) 
proportional to $\partial_\eta$. We write the energy defined by 
the Killing fields $\nu\partial_\eta$ as $E_{\rm R}$.

An observer fixed in the Rindler frame (with the trajectory 
where $x$ and $y$ are fixed, and $\xi=\xi_0=[{\rm constant}]$) 
is called a Rindler observer. 
The 4-velocity and 4-acceleration of the observer are 
$\tilde{u}^\mu=(-X^\alpha X_\alpha)^{-1/2}X^\mu$ and 
$\tilde{g}^\mu=\tilde{u}^\alpha\nabla_\alpha \tilde{u}^\mu$, and 
the worldline is expressed as a hyperbola, $z^2-t^2={\xi_0}^2$, in 
the inertial frame. Since the Rindler observer moves in a fixed direction 
with constant acceleration 
$\sqrt{\tilde{g}^\mu \tilde{g}_\mu}={\xi_0}^{-1}$, 
the Rindler frame is interpreted as a uniformly accelerated frame. 
The region  $h^+[t-z=0\  (\eta\rightarrow\infty,\ \xi\rightarrow 0)]$ 
is interpreted as the future event horizon of the Rindler frame, because 
the Rindler observer cannot see events beyond the region $h^+$. 
The region $h^-[t+z=0\ (\eta\rightarrow-\infty,\ \xi\rightarrow 0) ]$ 
is also interpreted as the past event horizon (see Fig. \ref{hyperps}).

If the emitted part and bound part satisfy conditions 2 and 3, 
we can evaluate the radiation power of the charge by using Eq. 
(\ref{tokkakari}). Then, let us calculate 
$E_{{\rm R}[\sigma(\eta\rightarrow\infty)]{\rm ret}}$. 
First we split the energy-momentum tensor ${T_{\rm ret}}^{\mu\nu}$ 
into the part ${T_{\I,\I}}^{\mu\nu}$, which is independent of  
$a^\mu$, the part ${T_{\I,\II}}^{\mu\nu}$, which is linear in 
$a^\mu$, and the part ${T_{\II,\II}}^{\mu\nu}$, which is quadratic in 
$a^\mu$ $[$ see Eqs. (2.7a) -- (2.7c) in Ref.~\cite{Te70} or see 
Eq.~(\ref{separate}) in this paper$]$, and evaluate the contributions 
to  the parts of $E_{{\rm R}[\sigma(\eta\rightarrow\infty)]{\rm ret}}$  
due to  ${T_{\I,\I}}^{\mu\nu}$, ${T_{\I,\II}}^{\mu\nu}$ and 
${T_{\II,\II}}^{\mu\nu}$, respectively. Explicit calculations 
are given in Appendices \ref{appendix} and \ref{new}, 
and we note here only the results.

The contribution independent of $a^\mu$ is
\begin{eqnarray}
\lim_{\eta\rightarrow\infty}E_{{\rm R}[\sigma(\eta)]\I,\I}&=&
-\lim_{\eta\rightarrow\infty}
\int_{\sigma(\eta)}n_\mu X_\nu {T_{\I,\I}}^{\mu\nu}\te
\nonumber\\
&=&\frac{2e^2}{3}
[v_\mu(\sqrt{g^\beta g_\beta}u^\mu+g^\mu)]^2(-\X^\alpha v_\alpha)d\tau,
\label{formI_I}
\end{eqnarray}
the contribution linear in $a^\mu$ is
\begin{eqnarray}
\lim_{\eta\rightarrow\infty}E_{{\rm R}[\sigma(\eta)]\I,\II}&=&
-\lim_{\eta\rightarrow\infty}\int_{\sigma(\eta)}n_\mu X_\nu 
{T_{\I,\II}}^{\mu\nu}\te
\nonumber\\
&=&
-\frac{4e^2}{3}a_\mu(\sqrt{g^\beta g_\beta}u^\mu +g^\mu)
(-\X^\alpha v_\alpha)d\tau,
\label{formI_II}
\end{eqnarray}
and the contribution quadratic in $a^\mu$ is
\begin{eqnarray}
\lim_{\eta\rightarrow\infty}E_{{\rm R}[\sigma(\eta)]\II,\II}&=&
-\lim_{\eta\rightarrow\infty}
\int_{\sigma(\eta)}n_\mu X_\nu {T_{\II,\II}}^{\mu\nu}\te
\nonumber\\
&=&\frac{2e^2}{3}a^\mu a_\mu(-\X^\alpha v_\alpha)d\tau
\label{formII_II},
\end{eqnarray}
where $u^\mu$, $g^\mu$ and $\X^\mu$ are the values of 
$\tilde{u}^\mu$, $\tilde{g}^\mu$ and $X^\mu$ at the retarded 
point $\bar{p}$, and $d\tau$ is the proper time interval of 
the charge between $\bar{p}$ and $\bar{p}'$. We also note that 
$-\X^\alpha v_\alpha>0$.

We would like to point out here that the above results explicitly 
show that the splitting of the energy-momentum tensor into the 
parts $\I$ and $\II$ performed by Teitelboim does not satisfy 
conditions 1 and 3 in the Rindler frame. On the other hand,  
condition 2 remains valid,  because part $\II$ is produced 
by the field $f^{\mu\nu}$ with $\lp^\mu=a^\mu$.

By summing up these contributions, we obtain the energy of the total 
electromagnetic field in the region 
$\sigma(\eta\rightarrow\infty)$ as
\begin{eqnarray}
\lim_{\eta\rightarrow\infty}E_{{\rm R}[\sigma(\eta)]{\rm ret}}&=&
\lim_{\eta\rightarrow\infty}[E_{{\rm R}[\sigma(\eta)]\I,\I}
+E_{{\rm R}[\sigma(\eta)]\I,\II}+E_{{\rm R}[\sigma(\eta)]\II,\II}]
\nonumber\\
&=&\frac{2e^2}{3}\alpha^{\mu}\alpha_\mu(-\X^\beta v_\beta)d\tau.
\label{formula}
\end{eqnarray}
Here
\begin{eqnarray}
\alpha^\mu&=&
h^\mu_{\ \nu}[a^\nu-w^\nu],\nonumber\\
w^\mu&=&(g^\alpha g_\alpha)^{1/2}u^\mu+g^\mu,
\end{eqnarray}
where $w^\mu$ is a null vector:
\begin{eqnarray}
w^\mu w_\mu&=&0.
\end{eqnarray}

We can write $\alpha^\mu=a^\mu-\we^\mu$ with the notation 
$\we^\mu=h^\mu_{\ \nu}w^\nu$. At the instant that the charge 
is at rest in the Rindler frame ($v^\mu=u^\mu$), we find 
$\alpha^\mu=a^\mu-g^\mu$. Hence $\alpha^\mu$ can be interpreted as 
the acceleration of the charge relative to the Rindler frame.

Equation~(\ref{formula}) resembles the Larmor formula 
in inertial frames,
\begin{eqnarray}
\lim_{t\rightarrow\infty}E_{{\rm M}[\sigma(t)]{\rm ret}}&=&
\frac{2e^2}{3}a^\mu a_\mu(-\bar{\partial_t}^\alpha v_\alpha)d\tau,
\label{Larmor-formula}
\end{eqnarray}
where $E_{\rm M}$ is the energy defined with respect to the Killing 
vector field $\partial_t$ of the inertial frame, $\sigma(t)$ is the 
intersection of simultaneous surface in the inertial frame with  
$\Delta L$. By replacing $\partial_t$ and $a^\mu$ with $X^\mu$ and 
$\alpha^\mu$ in Eq.~(\ref{Larmor-formula}), we again obtain the  
expression in Eq.~(\ref{formula}).
In analogy to the fact that the radiation fields $\II$ are linear in 
$a^\mu$ and the bound fields $\I$ are independent 
of $a^\mu$ in an inertial frame, 
we split the retarded fields into a part that is linear in 
$\alpha^\mu$ and a part that is independent of $\alpha^\mu$: 
\begin{eqnarray}
{F_{\rm ret}}^{\mu\nu}&=&{F_{\ \tI}}^{\mu\nu}+{F_{\ \tII}}^{\mu\nu},
\nonumber\\
{F_{\ \tI}}^{\mu\nu}&=&
\frac{e}{R^3}(1+\we^\alpha R_\alpha)[v^\mu R^\nu-R^\mu v^\nu]
+\frac{e}{R^2}[\we^\mu R^\nu-R^\mu\we^\nu],
\nonumber\\
{F_{\ \tII}}^{\mu\nu}&=&
\frac{e}{R^3}(\alpha^\alpha R_\alpha)[v^\mu R^\nu-R^\mu v^\nu]
+\frac{e}{R^2}[\alpha^\mu R^\nu-R^\mu\alpha^\nu].
\label{definition}
\end{eqnarray}
This splitting of the field causes a splitting of the energy-momentum 
tensor into two parts:
\begin{eqnarray}
{T_{\rm ret}}^{\mu\nu}&=&{T_{\ \tI}}^{\mu\nu}+{T_{\ \tII}}^{\mu\nu},
\nonumber
\end{eqnarray}

\begin{eqnarray}
{T_{\ \tI}}^{\mu\nu}&=&
\frac{1}{4\pi}\left[
{F_{\ \tI}}^{\mu\alpha}{F_{\ \tI}}_\alpha^{\ \nu}
+\frac{1}{4}g^{\mu\nu}
{F_{\ \tI}}_{\alpha\beta}{F_{\ \tI}}^{\alpha\beta}
\right.\nonumber\\
& &\hspace{0.8cm}\left.
+{F_{\ \tI}}^{\mu\alpha}{F_{\ \tII}}_\alpha^{\ \nu}
+{F_{\ \tII}}^{\mu\alpha}{F_{\ \tI}}_\alpha^{\ \nu}
+\frac{1}{2}g^{\mu\nu}{F_{\ \tI}}_{\alpha\beta}{F_{\ \tII}}^{\alpha\beta}
\right],
\nonumber\\
{T_{\ \tII}}^{\mu\nu}&=&
\frac{1}{4\pi}\left[
{F_{\ \tII}}^{\mu\alpha}{F_{\ \tII}}_\alpha^{\ \nu}+
\frac{1}{4}g^{\mu\nu}{F_{\ \tII}}_{\alpha\beta}{F_{\ \tII}}^{\alpha\beta}
\right].\label{deftII}
\end{eqnarray}

Since ${F_{\ \tII}}^{\mu\nu}$ corresponds to the field $f^{\mu\nu}$ with 
$\lambda^\mu=a^\mu-w^\mu$, and since $a^\mu-w^\mu$ satisfies  
Eq.~(\ref{example}),  
${F_{\ \tII}}^{\mu\nu}$ can be regarded as the radiation fields  that 
satisfy the conditions 1 and 2. By using the results of previous 
subsection, we find that the following equations are valid off the 
world line of the charge:
\begin{eqnarray}
\nabla_{[\lambda}{F_{\ \tII}}_{\mu\nu]}&=&0,
\label{four_potential}\\
\nabla_\mu {F_{\ \tII}}^{\mu\nu}&=&
-\frac{2e}{R^4}(\alpha^\mu R_\mu)R^\nu,
\label{Rnatural}\\
\nabla_\mu{T_{\ \tII}}^{\mu\nu}&=&0,\\
{T_{\ \tII}}^{\mu\nu}&=&
\frac{e^2}{4\pi R^6}
[(\alpha^\beta R_\beta)^2-R^2(\alpha^\beta\alpha_\beta)]R^\mu R^\nu.
\label{gutai}
\end{eqnarray}

From the equations concerning the total retarded fields and their 
energy-momentum tensor $\nabla_{[\lambda}{F_{\rm ret}}_{\mu\nu]}=0$,
$\nabla_\mu {F_{\rm ret}}^{\mu\nu}=0$ and 
$\nabla_\mu{T_{\rm ret}}^{\mu\nu}=0$, and from the equations 
concerning part $\tII$ above, we find that ${F_{\ \tI}}^{\mu\nu}$ 
and ${T_{\ \tI}}^{\mu\nu}$ satisfy the following equations off the world 
line of the charge:
\begin{eqnarray}
\nabla_{[\lambda}{F_{\ \tI}}_{\mu\nu]}&=&0,\\
\nabla_\mu {F_{\ \tI}}^{\mu\nu}&=&
\frac{2e}{R^4}(\alpha^\mu R_\mu)R^\nu,\\
\nabla_\mu{T_{\ \tI}}^{\mu\nu}&=&0.
\end{eqnarray}

Now, from the equation
\begin{eqnarray}
\lim_{\eta\rightarrow\infty}E_{{\rm R}[\sigma(\eta)]\tI}
&=&
\lim_{\eta\rightarrow\infty}E_{{\rm R}[\sigma(\eta)]{\rm ret}}
-E_{{\rm R}[\sigma(\eta)]\tII}
\end{eqnarray}
and the equation
\begin{eqnarray}
E_{{\rm R}[\sigma(\eta)]\tII}
&=&
\frac{2e^2}{3}\alpha^\mu\alpha_\mu(-\X^\alpha v_\alpha)d\tau,
\label{tIIrad}
\end{eqnarray}
which is derived by using Eq.~(\ref{gutai}) and the calculation 
in the appendices (see the comment below Eq.~(\ref{last}) in 
Appenix \ref{new}),
we find that
\begin{eqnarray}
\lim_{\eta\rightarrow\infty}E_{{\rm R}[\sigma(\eta)]\tI}&=&0.
\end{eqnarray}
Therefore, part $\tI$ of the energy-momentum tensor is regarded 
as the bound part, which satisfies condition 3.

In conclusion,  we find that part $\tI$ and part $\tII$, 
which are defined in Eqs. (\ref{definition}) and (\ref{deftII}), 
can be considered as the bound part and the emitted part in the Rindler 
frame, respectively, in the sense that these parts satisfy the 
conditions 1, 2 and 3.

\section{Physical implications of our identification}
\label{motion}

In this section, we investigate 
the implications of our identification of the emitted part and bound 
part in the Rindler frame given in the previous section.
Although we have distinguished ${\bar{x}}^\mu$ from $x^\mu$ and 
$\X^\mu$ from $X^\mu$ in the previous section and Appendices 
\ref{appendix} and \ref{new}, we do 
not make a distinction between these notations in this section, 
because we deal with quantities only on the world line of the 
charge here.

\subsection{Trajectory of a point charge with no radiation}

In our identification of radiation fields, the charge emits  
radiation neither in the case that the charge is fixed to the Rindler 
frame, nor in the case that $\alpha^\mu=0$. Thus, in this 
subsection, we investigate the trajectory of the charge which 
satisfies the equation $\alpha^\mu=0$.

From the 4-velocity of the Rindler observer in the Rindler frame,
\begin{eqnarray}
(u^\eta, u^x, u^y, u^\xi)&=&
(\xi^{-1}, 0, 0, 0),
\label{uR}
\end{eqnarray}
and the non-zero components of the Christoffel symbol in the Rindler frame, 
\begin{eqnarray}
\Gamma^\eta_{\xi\eta}
&=&\Gamma^\eta_{\eta\xi}=\xi^{-1},\ \ \ \ \Gamma^\xi_{\eta\eta}=\xi,
\label{Gamma}
\end{eqnarray}
we obtain the 4-acceleration $g^\mu=u^\alpha\nabla_\alpha u^\mu$ in 
the Rindler frame,
\begin{eqnarray}
(g^\eta, g^x, g^y, g^\xi)&=&
(0, 0, 0, \xi^{-1}).
\label{gR}
\end{eqnarray}
Taking covariant derivatives of Eqs. (\ref{uR}) and (\ref{gR}) gives 
\begin{eqnarray}
v^\mu\nabla_\mu g^\nu&=&
-(v\cdot g)g^\nu-(g\cdot g) (v\cdot u) u^\nu,
\label{difg}\\
v^\mu\nabla_\mu u^\nu&=&-(u\cdot v)g^\nu,
\label{difu}
\end{eqnarray}
where we use the notation $(p\cdot q)=p^\alpha q_\alpha$.
Equations (\ref{difg}) and (\ref{difu}) further give
\begin{eqnarray}
v^\mu\nabla_\mu\sqrt{g\cdot g}&=&
-(v\cdot g)\sqrt{g\cdot g},
\label{grav}\\
v^\mu\nabla_\mu w^\nu&=&v^\mu\nabla_\mu[\sqrt{g\cdot g}u^\nu+g^\nu]
=-(v\cdot w)w^\nu.
\label{difw}
\end{eqnarray}
From Eq.~(\ref{difw}) and  
$v^\mu\nabla_\mu h^\alpha_{\ \beta}=a^\alpha v_\beta+v^\alpha a_\beta$, 
we find
\begin{eqnarray}
v^\mu\nabla_\mu \we^\nu&=&
(v\cdot w)\alpha^\nu+v^\nu(a\cdot w).
\label{wedif}
\end{eqnarray}

According to the Lorentz-Dirac equations, the radiation reaction force 
acting 
on the charge is given as
\begin{eqnarray}
\Gamma^\mu&=&\frac{2e^2}{3}[v^\alpha\nabla_\alpha a^\mu-v^\mu(a\cdot a)].
\label{react}
\end{eqnarray}
By using Eq.~(\ref{wedif}), we find that the radiation reaction 
force vanishes for a charge with motion such that 
$\alpha^\mu=0$, or $a^\mu=\we^\mu$:
\begin{eqnarray}
\Gamma^\mu=0.
\label{nonreaction}
\end{eqnarray}
We thus find the natural result that a charge that does not radiate 
in the Rindler frame also does not experience a radiation reaction force.  
The trajectory with $\Gamma^\mu=0$ constitutes uniformly accelerated 
motion.
$[$According to the discussions of Sections 5.3 and 6.11 in Ref.~\cite{Ro90}, uniformly 
accelerated motion is defined by the equation 
$v^\lambda\nabla_\lambda a^\mu=a^\lambda a_\lambda v^\mu$, which is 
equivalent to Eq.~(\ref{nonreaction}).$]$ Thus there is an 
inertial frame $(t'',x'',y'',z'')$ in which the trajectory 
$\Gamma^\mu=0$ represents hyperbolic motion $[$see Eq.~(6$\cdot$127) 
in Section 6.11 of Ref.~\cite{Ro90}$]$ given by
\begin{eqnarray}
t''&=&\kappa^{-1}\sinh (\kappa \tau)\nonumber\\
z''&=&\kappa^{-1}\cosh (\kappa \tau),
\label{hyper}
\end{eqnarray}
where $\kappa$ is the acceleration of the trajectory $\alpha^\mu=0$ 
and evaluated as
\begin{eqnarray}
\kappa&=&\sqrt{\we\cdot\we}=v^\eta-\xi^{-1}v^\xi.
\label{kappa}
\end{eqnarray}
We note that, for uniformly accelerated motion 
$v^\alpha\nabla_\alpha a^\mu=a^\alpha a_\alpha v^\mu$, it follows 
that $v^\alpha\nabla_\alpha(a^\mu a_\mu)=0$, so that $\kappa$ is a 
constant throughout the motion (see Section 6.11 in Ref.~\cite{Ro90}).

Let us investigate the equation $a^\mu=\we^\mu$ by means of 
the Rindler coordinates.
Substituting Eqs. (\ref{uR}) -- (\ref{gR}) into 
the equation $a^\mu=\we^\mu$, we find the $\eta$, $\xi$, $x$ and $y$ 
components of the equation as 
\begin{eqnarray}
\frac{dv^\eta}{d\tau}&=&
\xi^{-2}-\xi^{-1}v^\xi v^\eta-(v^\eta)^2,
\label{eqeta}\\
\frac{dv^\xi}{d\tau}&=&
\xi^{-1}+\xi^{-1}(v^\xi)^2-v^\eta v^\xi-\xi(v^\eta)^2,
\label{eqxi}\\
\frac{dv^x}{d\tau}&=&
(\xi^{-1}v^\xi-v^\eta)v^x,
\label{eqx}\\
\frac{dv^y}{d\tau}&=&
(\xi^{-1}v^\xi-v^\eta)v^y,
\label{eqy}
\end{eqnarray}
where $v^\mu=dx^\mu/d\tau$. In addition to these equations, we should 
consider the conditions $v^\mu v_\mu=-1$ and $v^\eta>0$.  (We also consider 
$\xi>0$.) The former condition ensures that $\tau$ is the proper time, and 
the latter condition ensures that the coordinate time $\eta$ and the 
proper time $\tau$ are oriented in the same direction.

By using Eqs. (\ref{eqeta}) and (\ref{eqxi}), we can explicitly confirm  
the above statement that acceleration $\kappa$ is conserved:
\begin{eqnarray}
\frac{d}{d\tau}(v^\eta-\xi^{-1}v^\xi)&=&0.
\label{eqkappa}
\end{eqnarray}
Equation~(\ref{eqeta}) can be rewritten in the form
\begin{eqnarray}
\frac{1}{2}\frac{d}{d\tau}[(\xi v^\eta)^2 -1]&=&
-v^\eta[(\xi v^\eta)^2 -1].
\label{damping}
\end{eqnarray}
We express the speed of the particle in the Rindler frame as
$V=\{(v^x)^2+(v^y)^2+(v^\xi)^2\}^{1/2}$. Then we can write 
$V^2=(\xi v^\eta)^2-1$ by noting $v^\mu v_\mu=-1$.
Substituting this expression into Eq.~(\ref{damping}), it follows that
\begin{eqnarray}
\frac{dV}{d\eta}&=&-V.
\label{Vdamp}
\end{eqnarray}

\begin{figure}
\epsfxsize=12.5cm
\centerline{\epsfbox{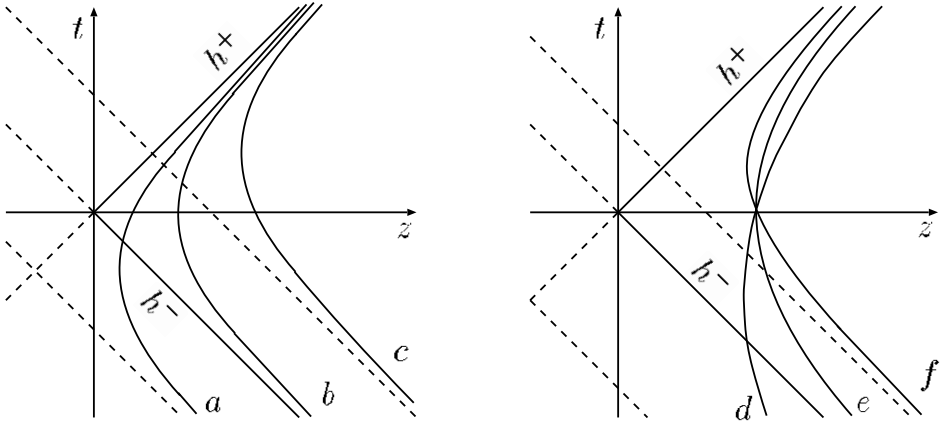}}
\caption{Trajectories given by the equation $\alpha^\mu=0$  are 
illustrated in the special case that $v^x=v^y=0$. Trajectories $a$, 
$b$ and $c$ experiencing the same acceleration $\kappa$ are asymptotically 
at rest in the Rindler frame with the same coordinate value 
$\xi=\kappa^{-1}$. Trajectories with velocities $v^\xi>0$,  $v^\xi=0$ 
and $v^\xi<0$ are illustrated as the curves $d$, $e$ and $f$. We find  
that the trajectory with $v^\xi>0$ intersects the past event horizon 
$h^-$.}
\label{hyperps}
\end{figure}

\hspace{-0.8cm}Then we find that $V\propto e^{-\eta}$ and that the speed $V$ decreases 
as the time $\eta$ is increased.

By integrating the set of Eqs. (\ref{eqkappa}), (\ref{damping}), 
(\ref{eqx}) and (\ref{eqy}) , we obtain the solution of equation
$a^\mu=\we^\mu$ (for details of the calculation, see 
Appendix \ref{details}):
\begin{eqnarray}
& &\sqrt{e^{2\eta}+A}=(\kappa\zeta)^{-1}e^{\kappa\tau}+B,
\label{soleta}\\
& &\xi=\zeta\left[
\{(\kappa\zeta)^{-1}+Be^{-\kappa\tau}\}^2-Ae^{-2\kappa\tau}
\right]^{1/2}
\label{solxi},\\
& &x=\beta_x e^{-\kappa\tau}+x_f,
\label{solx}\\
& &y=\beta_y e^{-\kappa\tau}+y_f,
\label{soly}
\end{eqnarray}
where the constants of integration $\kappa$, $\zeta$, $A$, $B$,  $\beta_x$, 
$x_f$, $\beta_y$ and $y_f$ satisfy following conditions:

\begin{eqnarray}
& &\kappa>0,\ \ \ \zeta>0\nonumber\\
& &B^2-A\zeta^2+{\beta_x}^2+{\beta_y}^2=0
\label{proper_time}.
\end{eqnarray}
By taking the limit $\tau\rightarrow\infty$ of this solution, 
we find that
\begin{eqnarray}
\eta&\rightarrow&\kappa\tau-\frac{1}{2}\ln(\kappa\zeta),
\\
(x,y,\xi)&\rightarrow&
(x_f,y_f,\kappa^{-1}).
\label{rest}
\end{eqnarray}
We thus see that the particle approaches a state of being at rest in the 
Rindler frame in the infinite future.

In conclusion $[$Eqs. (\ref{nonreaction}), (\ref{Vdamp}) and 
(\ref{rest})$]$, we find that the velocity of a particle that obeys 
the equation $a^\mu=\we^\mu$ 
gradually  decreases relative to the Rindler frame, and, in 
the limit $\eta\rightarrow\infty$, the particle is at rest in the Rindler 
frame. Also we find that the charge does not experience a  
radiation reaction force, which implies that there is an inertial 
frame in which the particle 
exhibits the hyperbolic motion described by Eq.~(\ref{hyper}). 
We illustrate the 
trajectories with $a^\mu=\we^\mu$ in the special case that $v^x=v^y=0$ 
in Fig. \ref{hyperps}.

\subsection{Bound electromagnetic energy in the Rindler frame}
\label{subbound}

In Teitelboim's original paper~\cite{Te70},  the bound character 
of part $\I$ is strongly confirmed by showing that the 4-momentum 
of part $\I$ is a state function of the charge.
In that paper, the 4-momentum of the part $\I$ is defined as  
\begin{eqnarray}
{P_{\I}}^\mu(t)&=&\int_{\Sigma(t)}{T_{\I}}^{\nu\mu}n_\nu\te,
\label{defPI}
\end{eqnarray}
expressed in the inertial frame, where  $\Sigma(t)$ is a space-like 
plane in Minkowski spacetime, and $n^\nu$ is a future directed unit 
vector normal to the plane $\Sigma(t)$. 
Since ${T_{\I}}^{\nu\mu}n_\nu$ is infinite on the world line of the 
charge, one integrates it over $\Sigma(t)$, except inside the sphere 
of small radius $\epsilon$ that surrounds the charge.

In the 
original paper~\cite{Te70},  $\Sigma(t)$ is set orthogonal to the 
world line of the charge,\footnote{We note here that 
if $\Sigma(t)$ is not orthogonal 
to the world line, 
we get another result $[$see Eqs. (7.9) and (C.9) in the review 
written by Teitelboim et al.~\cite{TV80}$]$.  Even in this case, the 
bound character of part 
$\I$, the property that the 4-momentum of part $\I$ is a state 
function of the charge, is confirmed.} 
 and 
the result is $[$ Eq.~(3.20) in Ref.~\cite{Te70}$]$
\begin{eqnarray}
{P_{\I}}^\mu&=&\frac{e^2}{2\epsilon}v^\mu-\frac{2e^2}{3} a^\mu.
\label{boundE}
\end{eqnarray}
Although  ${T_{\I}}^{\nu\mu}$ 
in Eq.~(\ref{defPI}) is a retarded function, and thus  ${P_{\I}}^\mu$ 
is defined by the integration over the entire history of the charge, 
the total integral depends only on  the instantaneous state of the 
charge $[v^\mu$ and $a^\mu$ in Eq.~(\ref{boundE})$]$.
This noteworthy result implies that we can regard part $\I$ as 
bound to charge~\cite{Te70}.

To strongly confirm the bound character of part $\tI$ in 
the Rindler frame, we should evaluate the total energy of $\tI$ in the Rindler 
frame. Although we do not explicitly evaluate the total energy here, we 
make a rough estimation concerning the interaction of the charge with 
part $\tI$ in terms of the energy defined in the Rindler frame, by 
using the Lorentz-Dirac equations.

The energy of a particle in a stationary frame generated by the Killing 
vector field $X^\mu$ may be written as
\begin{eqnarray}
E_{\rm Kp}&=&-X^\mu v_\mu,
\end{eqnarray}
where we assume a particle with unit mass. By using the Killing 
equation
$\nabla_\mu X_\nu+\nabla_\nu X_\mu=0$, we obtain the rate of change of the 
energy of the particle per unit proper time as
\begin{eqnarray}
v^\alpha\nabla_\alpha E_{\rm Kp}&=&-X^\mu (v^\alpha\nabla_\alpha v_\mu).
\label{dE}
\end{eqnarray}
Of course, for a free particle ($v^\lambda\nabla_\lambda v_\mu=0$),  
$E_{\rm Kp}$ is a constant.

From the Lorentz-Dirac equations 
$v^\alpha\nabla_\alpha v^\mu={F_{\rm ext}}^\mu+\Gamma^\mu$ and 
Eq.~(\ref{dE}), 
we can infer that the energy 
$\Delta W_{\rm R}$ provided by the moving charge for the electromagnetic 
field per proper time is $X^\mu \Gamma_\mu$. Substituting 
$a^\mu=\alpha^\mu+\we^\mu$ into Eq.~(\ref{react}), and applying 
Eq.~(\ref{wedif}), we obtain
\begin{eqnarray}
\frac{3}{2e^2}\Gamma^\nu&=&
v^\mu\nabla_\mu\alpha^\nu+(v\cdot w)\alpha^\nu-v^\nu(\alpha\cdot w)
-v^\nu(\alpha\cdot\alpha).
\label{reacR}
\end{eqnarray}
By noting $X^\mu\propto(g\cdot g)^{-1/2}u^\mu$, and using Eqs. 
(\ref{difu}) and (\ref{grav}), we get, for the Killing vector 
field of the Rindler frame,
\begin{eqnarray}
v^\mu\nabla_\mu X^\nu&=&
(v\cdot g)X^\nu-(X\cdot v)g^\nu\nonumber\\
&=&(v\cdot w)X^\nu-(X\cdot v)w^\nu.
\label{difX}
\end{eqnarray}
From Eqs. (\ref{reacR}) and (\ref{difX}), it follows that
\begin{eqnarray}
\Delta W_{\rm R}&=&X^\mu \Gamma_\mu\nonumber\\
&=&
\frac{2e^2}{3}v^\mu\nabla_\mu(X\cdot\alpha)
-\frac{2e^2}{3}(X^\alpha v_\alpha)\alpha^\mu\alpha_\mu.
\end{eqnarray}
According to Eq.~(\ref{tIIrad}),
the second term on the right-hand side of this equation,
$-(2e^2/3)(X^\alpha v_\alpha)\alpha^\mu\alpha_\mu$,  
is interpreted as the 
contribution to the  emitted part $\tII$, so that we can infer 
that the contribution to the bound part $\tI$ is 
\begin{eqnarray}
\frac{2e^2}{3}v^\mu\nabla_\mu(X\cdot\alpha)&=&
\frac{d}{d\tau}
\left[
\frac{2e^2}{3}X^\mu\alpha_\mu
\right].
\label{dxa}
\end{eqnarray}
Compared with the calculation concerning part $\I$ performed 
by Teitelboim, this result is analogous to Eq.~(3.18) in 
Ref.~\cite{Te70}. 
The fact that Eq.~(\ref{dxa}) has an integrable form implies that 
the Rindler energy of part $\tI$ would include the contribution  
$(2e^2/3)X^\mu\alpha_\mu$, which is analogous to the second term 
(Schott term) on the right-hand side of Eq.~(\ref{boundE}).

\section{Discussion}
\label{discussion}

In the Rindler frame, we have identified the emitted and bound 
parts of the energy-momentum tensor of the retarded fields, in 
accordance with the three conditions introduced in \S~\ref{Kgeneral}.
Equations derived here in the Rindler frame are regarded as natural 
extensions of equations in the inertial frame derived by Teitelboim 
(with the replacement of $\X$ for $\bar{\partial}_t$ and $\alpha^\mu$ 
for $a^\mu$).

In application of our work
to general accelerated frames with Killing fields other than the Rindler 
frame, condition 2 would remain valid, because this condition is 
crucial to define radiation. Conditions 1 and 3, however, 
may be replaced with other conditions.
To examine the applicability
of condition 1 in general accelerated frames, it might be 
effective to examine whether the Poynting flux generated by a charge 
fixed in the frame is zero, as seen by an observer fixed in the same 
frame.
Also, as mentioned in \S~\ref{subbound},   to identify the bound 
part of the Maxwell tensor, it is important to determine whether the total 
energy of this part is a state function of the charge. This requirement,  
which provides a physical picture for the definition of the bound part, 
seems to be more essential than the formal requirment of condition 3.

In our result, a charge in the Rindler frame radiates neither      
in the case that the charge is fixed in the Rindler frame, nor 
in the case that the charge follows the trajectory $\alpha^\mu=0$. 
There are studies concerning classical radiation in the Rindler frame, 
where the non-existence of radiation from the charge fixed in 
the Rindler frame is discussed, but the non-existence of 
radiation from the charge with other trajectories has not been 
discussed (or, at least, the author does not know of such studies).
Therefore, this might provide a new point of view in the study of 
radiation in the Rindler frame.

There is an old paradox of classical electrodynamics concerning the 
radiation from a charge with uniformly accelerated 
motion~\cite{FR60,Bo80}.   We now apply our result in 
the Rindler frame to this problem.
It seems that this  paradox has been discussed in subtly distinct 
situations or points of view in the literature. Here we restrict our 
attention to the following problem. Let us compare the situation in 
which a charge exhibits uniformly accelerated motion 
with the situation in which a charge is fixed in a static 
gravitational field. These two 
situations might be distinguished by examining the radiation from 
the charge, because, in the former case, the charge radiates according 
to the Larmor formula, and in the latter case it does not radiate 
due to the energy conservation. However, this conclusion seems to 
contradict the principle of equivalence, which asserts that one cannot 
distinguish an accelerated frame from a gravitational field.
This paradox may be resolved by noting that the concept (or detection) 
of radiation depends on the observer. In the above described situation, 
radiation in 
the gravitational field is defined  with respect to the observer  
supported in that field. Therefore one should compare the radiation 
defined by this observer with the radiation defined by the observer 
fixed in the accelerated frame. In this case, the Larmor formula is 
not applicable, because this formula defines radiation with respect 
to an inertial observer. As mentioned above, in the literature, it 
has been asserted that an observer fixed in the Rindler frame 
does not detect radiation from a charge fixed in the same  
frame because of the vanishing of the Poynting flux~\cite{Bo80}.

The following four {\it gedanken experiments} are often disscussed 
in the context of the observer dependence of the concept of radiation.
Let us consider whether the observer detects radiation from the charge 
in the following four situations:\footnote{Our conclusion here is 
equivalent to that obtained by Fugmann and Kretzschmar in 
Ref.~\cite{FK91} (see the footnote in the Introduction and 
Appendix~\ref{comparison} ).}\\
\begin{enumerate}

\item[A.]{\it a charge fixed in an inertial frame, with an observer 
fixed in the same frame},

\item[B.]{\it a charge in uniform acceleration, with an observer 
fixed in an inertial frame},

\item[C.]{\it a charge in inertial motion, with an observer fixed in a 
uniformly accelerated frame},

\item[D.]{\it a charge fixed in a uniformly accelerated frame, with an 
observer fixed in the same frame}.\\

\end{enumerate}

In cases A and B, we can use the definition of 
radiation given by 
Rohrlich and Teitelboim (Larmor formula), with the result that 
the radiation is detected in case B, but not detected 
in case A. 
In cases C and D, where the observer experiences 
uniformly accelerated 
motion, it would be appropriate to use the definition of radiation  
in the Rindler frame $[$Eq.~(\ref{formula}) or Eq.~(\ref{tIIrad})$]$. 
Then, it is concluded that the observer does not detect radiation 
in case D, because our identification of radiation satisfies  
condition 1, but does detect radiation in case C, because $\alpha^\mu\ne 0$.

Our work in the Rindler frame provides a picture in which the concept of 
radiation depends on the observer. We here note that
there is an analogous situation 
in quantum field theory,\footnote{See Section 8 in Ref.~\cite{FK91}.} 
which is called `Hawking effect' or `Unruh effect'~\cite{BD82,Ta86}.   
This effect consists of the prediction in quantum field theory that
a uniformly accelerated observer in a vacuum  perceives a thermal 
bath of particles with temperature proportional to his 
acceleration. This prediction provides the physical interpretation 
that an observer experiencing inertial motion and an observer 
experiencing uniform 
acceleration do not agree on the number of particles observed. 
Therefore we find that, in the quantum case, the concept of particles 
depends on the observer, analogous to the observer dependence of 
the concept of radiation in classical 
theory.\footnote{See Ref.~\cite{PV99} and the references therein 
for discussions about this point of view.} It might be interesting 
to study classical radiation in accelerated frames and 
general curved spacetimes to investigate the connection to 
quantum theory~\cite{HM93}.

\section*{Acknowledgements}
I wish to thank Professor Tetsuya Hara for his encouragement and 
for helpful advice and discussion.
\appendix
\section{Radiation formula in the Rindler frame} 
\label{appendix}

In this appendix, we derive 
Eqs. (\ref{formI_I}) -- (\ref{formII_II}), 
to obtain the radiation formula (\ref{formula}) in the Rindler frame.
First, we calculate the Rindler energy ${E_{\rm R}}^\flat$ with the the 
Killing vector field $X=\partial_\eta\equiv X^\flat$. Then we obtain 
the Rindler energy $E_{\rm R}$ with the Killing vector field 
$X=\nu\partial_\eta$  by multiplying  ${E_{\rm R}}^\flat$ by $\nu$.

We introduce cylindrical coordinates $(\xi,\rho,\varphi)$ in a 
simultaneous plane  $\eta=[{\rm constant}]$ in the Rindler frame.
$\rho$ and $\varphi$ are defined as
\begin{eqnarray}
x-\bx&=&\rho \cos\varphi,
\label{pol1}\\
y-\by&=&\rho \sin\varphi.
\label{pol2}
\end{eqnarray}
Here we assume that $\bx$ and $\by$ are the coordinates of the 
point $\bar{p}$ fixed on the world line of the charge, at which 
the radiation we evaluate is  emitted. 
With the notation used in Eqs. (\ref{pol1}) and (\ref{pol2}), $x$ and $y$ 
are defined throughout the region $\eta=[{\rm constant}]$, and $\bx$ 
and $\by$ are not retarded functions of $x$ and $y$. 
The volume element in the region $\eta=[{\rm constant}]$ is expressed 
as $\te=\rho d\rho d\phi d\xi$.

In our cylindrical coordinates, ${E_{\rm R}}^\flat$ is written
\begin{eqnarray}
\lim_{\eta\rightarrow\infty}E_{{\rm R}[\sigma(\eta)]}^{\ \flat}
&=&-\lim_{\eta\rightarrow\infty}
\int_{\sigma(\eta)}n_\mu {X^\flat}_\nu T^{\mu\nu}\te
\nonumber\\
&=&-\lim_{\eta\rightarrow\infty}
\int_0^{2\pi}d\varphi\int_{\xi_{\rm min}}^{\xi_{\rm max}}d\xi
|\rho d\rho|\xi^3 T^{\eta\eta},
\label{ERI}
\end{eqnarray}
where $\xi_{\rm min}$ and $\xi_{\rm max}$ are, respectively, 
the minimum value and the  maximum value of $\xi$ in the region 
$\sigma(\eta)$. Let us rewrite
$\rho d\rho$ in Eq.~(\ref{ERI}) with respect to the proper time 
$d\tau$ between the point $\bar{p}$ (with coordinate values 
$\bar{x}^\mu$) and the point $\bar{p}'$ (with coordinate values 
$\bar{x}'^{\mu}$). Let us assume that the points $x^\mu$ and 
$x'^\mu$ are, respectively, the points in the future light cones 
with apexes $\bar{x}^\mu$ and $\bar{x}'^{\mu}$, where 
$\eta=\eta',\ \ \xi=\xi',\ \ \varphi=\varphi'$. With the definitions 
$R^\mu=x^\mu-\bar{x}^\mu$ and $R'^\mu=x'^\mu-\bar{x}'^\mu$, 
it follows that
\begin{eqnarray}
R^\mu R_\mu&=&
\xi^2+\bi^2-2\xi\bi\cosh(\eta-\be)+\rho^2,
\label{RR}
\\
R'^\mu R'_\mu&=&
\xi^2+\bi'^2-2\xi\bi'\cosh(\eta-\be')+(x'-\bar{x}')^2+(y'-\bar{y}')^2.
\label{R'R'}
\end{eqnarray}
Substituting $\bar{x}'^\mu=\bar{x}^\mu+v^\mu d\tau$ into 
Eq.~(\ref{R'R'}),
and ignoring contributions of higher order in $d\tau$, 
we obtain
\begin{eqnarray}
R'^\mu R'_\mu&=&R^\mu R_\mu+\rho'^2-\rho^2+2Rd\tau,
\label{seiri}
\end{eqnarray}
where we have used Eq.~(\ref{RR}) and 
\begin{eqnarray}
\hspace{-1cm}R&=&-R^\mu v_\mu\nonumber\\
\hspace{-1cm}
&=&[\bi-\xi \cosh(\eta-\bar{\eta})]v^\xi
+\xi\bar{\xi}\sinh(\eta-\bar{\eta})v^\eta
-(x-\bar{x})v^x-(y-\bar{y})v^y.
\label{R}
\end{eqnarray}
By noting $R^\mu R_\mu=R'^\mu R'_\mu=0$ and 
$\rho'^2-\rho^2=2\rho d\rho$,
we find that Eq.~(\ref{seiri}) gives
\begin{eqnarray}
\rho d\rho=-Rd\tau.
\label{Rdt}
\end{eqnarray}

From Eq.~(\ref{RR}) and $R^\mu R_\mu=0$, it follows that
\begin{eqnarray}
(x-\bx)^2+(y-\by)^2+(\xi-\bi\ch)^2&=&\bi^2\sh^2,
\label{S}
\end{eqnarray}
where we have used the definitions
\begin{eqnarray}
\ch&=&\cosh(\eta-\be),\\
\sh&=&\sinh(\eta-\be).
\end{eqnarray}
According to Eq.~(\ref{S}),
we find that the region of integration in Eq.~(\ref{ERI}) is the 
spherical shell with radius $\bi\sh$ centered at the point 
$(\bx, \by, \bi\ch)$ in the space $(x, y,\xi)$. We refer to this 
spherical shell as $S$.  

Here we introduce the new cylindrical coordinates $(R^*,\mu,\theta)$ 
in the region 
$\eta=[{\rm constant}]$. $R^*$ is defined as
\begin{eqnarray}
R^*&=&
[\bi-\xi \ch]v^\xi+\xi\bar{\xi}\sh v^\eta-(x-\bar{x})v^x-(y-\bar{y})v^y,
\label{Rstar}
\end{eqnarray}
where $\be$, $\bi$, $\bx$, $\by$, $v^\eta$, $v^\xi$, $v^x$ and $v^y$ 
are evaluated at the fixed point $\bar{p}$, so that these are  
constant throughout the region $\eta=[{\rm constant}]$. (Of course, $R$ 
is also defined in the same form as in Eq.~(\ref{Rstar}), although, 
in this case, $\bx^\mu$ and $v^\mu$ are the retarded functions of 
$x^\mu$, so that these are not constants.) 
Now we introduce the vectors
\begin{eqnarray}
{\bf P}&=&
-(Y^x_{\ \ \mu}|_p v^\mu|_{\bar{p}}, Y^y_{\ \ \mu}|_p v^\mu|_{\bar{p}}, 
Y^\xi_{\ \ \mu}|_p v^\mu|_{\bar{p}}, )
\nonumber\\
&=&
-(v^x, v^y, v^\xi\ch- v^\eta\bi\sh )\nonumber\\
&\equiv&
-(v^x, v^y,P),
\label{P}\\
{\bf x}&=&(x-\bx, y-\by, \xi-\bi\ch)\nonumber\\
&\equiv&
({\bf x}_1,{\bf x}_2,{\bf x}_3),
\end{eqnarray}
where $Y^\mu_{\ \ \nu}$ represents a transformation from the inertial 
frame $(t,x,y,z)$ to the Rindler frame $(\eta, x, y, \xi)$ (the upper index 
of $Y^\mu_{\ \ \nu}$ representing the components in the Rindler frame, and 
the lower index representing the components in inertial frame), and 
$v^\mu$ in  
the first relation of Eq.~(\ref{P}) is expressed in the inertial frame. 
We also note that in Eq.~(\ref{P}),
 we transform the vector at the point $\bar{p}$ 
with the transformation given at the point $p$.
By using ${\bf P}$ and ${\bf x}$, $R^*$ can be written 
\begin{eqnarray}
R^*&=&
{\bf P}\cdot{\bf x}+\bi(-\ch P+v^\xi).
\end{eqnarray}
By noting the angle between ${\bf P}$ and ${\bf x}$, we find that the 
maximum of $R^*$ on the surface $S$ is 
${R^*}_{\rm max}=\bi[(-\ch P+v^\xi)+\sh |{\bf P}|]$, and the minimum 
is ${R^*}_{\rm min}=\bi[(-\ch P+v^\xi)-\sh |{\bf P}|]$. We also obtain 
${R^*}_{\rm max}{R^*}_{\rm min}=\bi^2\sh^2$ by using $v^\mu v_\mu=-1$

Next, we define the coordinates $\mu$ and $\theta$. Now we set $v^y=0$ 
without loss of generality, since the geometry of the Rindler frame is 
invariant with respect to the rotation within the $x$-$y$ plane. 
In this case, ${\bf P}$ is embeded in the ${\bf x}_1$-${\bf x}_3$ plane.  
We introduce the  cylindrical coordinates $(R^*,\mu,\theta)$ with $R^*$ 
increasing in the direction of ${\bf P}$, the radial coordinate $\mu$ 
and the angular coordinate $\theta$. The coordintes 
$({\bf x}_1,{\bf x}_2,{\bf x}_3)$ are transformed into the coordinates 
$(R^*,\mu,\theta)$ according to 
\begin{eqnarray}
\hspace{-0.8cm}
\left(
\begin{array}{c}
\mu\cos\theta\\
\mu\sin\theta\\
|{\bf P}|^{-1}(R^*+\bi\ch P-v^\xi \bi)
\end{array}
\right)
&=&
\left(
\begin{array}{ccc}
\cos\Theta&0&-\sin\Theta\\
0&1&0\\
\sin\Theta&0&\cos\Theta
\end{array}
\right)
\left(
\begin{array}{c}
{\bf x}_1\\
{\bf x}_2\\
{\bf x}_3
\end{array}
\right),
\label{Rmutheta}
\end{eqnarray}
where $\cos\Theta=-|{\bf P}|^{-1}P$ and $\sin\Theta=-|{\bf P}|^{-1}v^x$,
in which $\Theta$ is the angle between ${\bf P}$ and the ${\bf x}_3$ axis.

From the expression ${\bf x}\cdot{\bf x}=\bi^2\sh^2$ for $S$ and 
Eq.~(\ref{Rmutheta}), we find the relation between $\mu$ and $R^*$ 
in the region $S$: 
\begin{eqnarray}
\mu^2+|{\bf P}|^{-2}(R^*+\bi\ch P-v^\xi \bi)^2&=&\bi^2\sh^2.
\label{constraint}
\end{eqnarray}
By using Eqs. (\ref{Rmutheta}) and (\ref{constraint}), we obtain 
on the 2-dimensional surface $S$
\begin{eqnarray}
dR^*\wedge d\theta&=&|{\bf P}|d\xi\wedge d\varphi.
\end{eqnarray}
Substituting this relation and Eq.~(\ref{Rdt}) into Eq.~(\ref{ERI}), 
we obtain
\begin{eqnarray}
\lim_{\eta\rightarrow\infty}E_{{\rm R}[\sigma(\eta)]}^{\ \flat}&=&
-\lim_{\eta\rightarrow\infty}\frac{d\tau}{|{\bf P}|}
\int_{{R^*}_{\rm min}}^{{R^*}_{\rm max}}dR^*
\int_0^{2\pi}d\theta R^*\xi^3 T^{\eta\eta}.
\label{ERindler}
\end{eqnarray}

${T_{\rm ret}}^{\mu\nu}$ is split into a part that is independent 
of $a^\mu$, a part that is linear in $a^\mu$ and the part that is 
quadratic in $a^\mu$ $[$Eqs.~(2.7a) -- (2.7c) in Ref.~\cite{Te70}$]$:
\begin{eqnarray}
{T_{\rm ret}}^{\mu\nu}&=&
{T_{\I,\I}}^{\mu\nu}+{T_{\I,\II}}^{\mu\nu}+{T_{\II,\II}}^{\mu\nu},
\label{separate}\\
\frac{4\pi}{e^2}{T_{\I,\I}}^{\mu\nu}&=&
\frac{1}{R^6}
\left[
R^\mu R^\nu-R(v^\mu R^\nu+R^\mu v^\nu)-\frac{1}{2}\eta^{\mu\nu}R^2
\right],\\
\frac{4\pi}{e^2}{T_{\I,\II}}^{\mu\nu}&=&
\frac{1}{R^6}
[2(a\cdot R)R^\mu R^\nu
\nonumber\\
& &
-R(a\cdot R)(v^\mu R^\nu+R^\mu v^\nu)
-R^2(a^\mu R^\nu+R^\mu a^\nu)],\\
\frac{4\pi}{e^2}{T_{\II,\II}}^{\mu\nu}&=&
\frac{1}{R^6}
[(a\cdot R)^2-R^2(a\cdot a)]
R^\mu R^\nu.
\label{IIandII}
\end{eqnarray}

Let us evaluate the contribution to 
$E_{{\rm R}[\sigma(\eta)]}^{\ \flat}$ due to the part 
${T_{\I,\I}}^{\mu\nu}$. By using 
$Y^\eta_{\ \ \mu}|_p R^\mu=\xi^{-1}\bi\sh$ and 
$Y^\eta_{\ \ \mu}|_p v^\mu|_{\bar{p}}=
\xi^{-1}[\bi\ch v^\eta-\sh v^\xi]
\equiv\xi^{-1}q$, $\xi^3{T_{\I,\I}}^{\eta\eta}$ is written as
\begin{eqnarray}
\frac{4\pi}{e^2}\xi^3{T_{\I,\I}}^{\eta\eta}
&=&
\frac{\xi}{R^6}\left[
\bi^2\sh^2-2\bi\sh q R +\frac{1}{2}R^2
\right].
\label{BBe}
\end{eqnarray}
$[$In the following, we do not distinguish $R^*$ and $R$, because $R^*=R$ 
in the region of integration in Eq.~(\ref{ERindler}).$]$
$\xi$ in Eq.~(\ref{BBe}) is rewritten in $R$ and $\theta$ by using 
Eq.~(\ref{Rmutheta}) and by noting ${\bf x}_3=\xi-\bi\ch$:
\begin{eqnarray}
\p^2\xi&=&
-P(R-v^\xi\bi)+\bi\ch(\p^2-P^2)+v^x\p\mu\cos\theta.
\label{pxi}
\end{eqnarray}
Here $\mu$ is a function of $R$ with the dependence given 
in Eq.~(\ref{constraint}):
\begin{eqnarray}
\p^2\mu^2&=&
\p^2\bi^2\sh^2-(R+\bi\ch P-v^\xi\bi)^2.
\label{pmu}
\end{eqnarray}

Substituting Eqs. (\ref{BBe}) and (\ref{pxi}) into 
Eq.~(\ref{ERindler}) and integrating it with respect to $\theta$, we 
find that the contribution that depends on $\theta$ vanishes, and the 
terms of order $R^{-2}$, $R^{-3}$, $R^{-4}$ and $R^{-5}$ remain. 
We can integrate these terms 
by using the values ${R}_{\rm min}$, ${R}_{\rm max}$ and 
${R}_{\rm max}{R}_{\rm min}$ obtained above. In the limit 
$\eta\rightarrow\infty$, where  
$\sh=\ch$ and $P=-\p<0$, we have
\begin{eqnarray}
\lim_{\eta\rightarrow\infty}
\int_{{R}_{\rm min}}^{{R}_{\rm max}}\frac{dR}{R^2}
&=&
2\bi^{-1}\ch^{-1}(-P),
\label{R2}\\
\lim_{\eta\rightarrow\infty}
\int_{{R}_{\rm min}}^{{R}_{\rm max}}\frac{dR}{R^3}
&=&2\bi^{-2}\ch^{-2}P^2,
\label{R3}\\
\lim_{\eta\rightarrow\infty}
\int_{{R}_{\rm min}}^{{R}_{\rm max}}\frac{dR}{R^4}
&=&\frac{8}{3}\bi^{-3}\ch^{-3}(-P^3),
\label{R4}\\
\lim_{\eta\rightarrow\infty}
\int_{{R}_{\rm min}}^{{R}_{\rm max}}\frac{dR}{R^5}
&=&4\bi^{-4}\ch^{-4}P^4.
\label{R5}
\end{eqnarray}
Since $P$ is a quantity of order 1 in $\ch$, all of these 
integrations give values of order 0 in $\ch$.  
Let us now  consider the expansion of the integrand in 
Eq.~(\ref{ERindler}) with respect to $R$, and evaluate the order 
in $\ch$ of each coefficient of the expansion.
If the order of a coefficient is smaller than 1, the term 
with this coefficients does not contribute to integration in the 
limit $\eta\rightarrow\infty$.  By using this  property, 
we can reduce our effort in the calculation of Eq.~(\ref{ERindler})
to some extent.

Since the third term in Eq.~(\ref{BBe}) does not contribute to 
the integration in the limit $\eta\rightarrow\infty$, we can write 
\begin{eqnarray}
\hspace{-1cm}\lim_{\eta\rightarrow\infty}
E_{{\rm R}[\sigma(\eta)]\I,\I}^{\ \flat}&=&
-\lim_{\eta\rightarrow\infty}
\frac{d\tau}{|{\bf P}|}
\int_{{R^*}_{\rm min}}^{{R^*}_{\rm max}}dR^*
\int_0^{2\pi}d\theta 
\frac{e^2}{4\pi}\frac{\xi}{R^5}
[\bi^2\ch^2-2\bi\ch qR].
\label{RindlerE}
\end{eqnarray}
According to the consideration given above, we find that only terms in $\xi$ 
that are of order $\geq -1$ in $\ch$ contribute to the integration:
\begin{eqnarray}
\p^2\xi
&\sim&-PR+\bi[Pv^\xi+\ch (v^x)^2]+v^x\p\mu\cos\theta,
\label{xiRth}
\end{eqnarray}
where, on the right-hand side of this equation, we have ignored terms of 
sufficiently low order, which do not contribute to the 
integration (\ref{RindlerE}).

Substitution of Eq.~(\ref{xiRth}) into Eq.~(\ref{RindlerE}) for $\xi$ 
reduces the expression of integrand to terms homogeneous in $\ch$.
Therefore, we can omit $\ch$ in the following calculations without 
confusion.
 When $\eta\rightarrow\infty$, we can use $q=-P$. Furthermore, since 
there are many terms of the form 
$Pv^\eta\bi$, $(v^x)^2\ch$ and $Pv^\xi$ in calculations, we can  
make the calculation somewhat easier by setting $a=Pv^\eta\bi$ 
(where the quantity $a$ is not related to $a^\mu$), 
$b=(v^x)^2\ch$ and $c=Pv^\xi$ with the relations $a+b+c=-1$ and  
$P^2=c-a$ satisfied in the limit $\eta\rightarrow\infty$ (where 
$\ch$ is omitted).
We begin the integration in Eq.~(\ref{RindlerE}) by performing that
over $\theta$. Then, we
perform the integration over $R$ by using Eqs. (\ref{R3}) -- (\ref{R5}). 
This yields 
\begin{eqnarray}
\lim_{\eta\rightarrow\infty}
E_{{\rm R}[\sigma(\eta)]\I,\I}^{\ \flat}&=&
\lim_{\eta\rightarrow\infty}\frac{2e^2}{3}\frac{d\tau}{\bi}Pa
=
\frac{2e^2}{3}d\tau(\bi v^\eta-v^\xi)^2 v^\eta.
\end{eqnarray}

Next, let us derive the contribution of 
$E_{{\rm R}[\sigma(\eta)]}^{\ \flat}$
due to ${T_{\I,\II}}^{\mu\nu}$. We proceed by noting 
that only the terms in 
${T_{\I,\II}}^{\mu\nu}$ of order 4 in $\ch$ contribute to the 
integration in the limit $\eta\rightarrow\infty$.   
${T_{\I,\II}}^{\mu\nu}$ is arranged into components of the 
acceleration $a^\mu$ as follows:
\begin{eqnarray}
\frac{2\pi}{e^2}\frac{\xi^2R^6}{\bi}{T_{\I,\II}}^{\eta\eta}&\sim&
[-\bi R^2-\xi\bi(PR+\bi)]a^\eta+[R^2+(\xi-\bi)(PR+\bi)]a^\xi
\nonumber\\
& &+(PR+\bi)(x-\bx)a^x+(PR+\bi)(y-\by)a^y,
\label{seta}
\end{eqnarray}
where we have omitted $\ch$. By using Eq.~(\ref{xiRth}) 
and the relations
\begin{eqnarray}
\p^2\mu^2&\sim&
-2\bi\ch PR-\bi^2\sh^2,\\
\p(x-\bx)&\sim&
\bi\ch v^x+\p\mu\cos\theta,\\
\p(y-\by)&\sim&
\p\mu\sin\theta,
\end{eqnarray}
we can perform the integration of the contribution due to 
each component of 
$a^\mu$ in Eq.~(\ref{seta}), so that
\begin{eqnarray}
\lim_{\eta\rightarrow\infty}
E_{{\rm R}[\sigma(\eta)]\I,\II}^{\ \flat}&=&
\frac{2e^2 d\tau}{3P}
\left[(1-2a)a\bi a^\eta+(c+2ac-2a)a^\xi+(2a+1)Pv^x a^x
\right]\nonumber\\
&=&\frac{2e^2 d\tau}{3P}
\left[
(2a+1)P(a^\mu v_\mu)+2a(\bi a^\eta-a^\xi)
\right].
\label{I-II}
\end{eqnarray}
By noting $a^\mu v_\mu=0$, we obtain
\begin{eqnarray}
\lim_{\eta\rightarrow\infty}
E_{{\rm R}[\sigma(\eta)]\I,\II}^{\ \flat}&=&
\frac{4e^2}{3}d\tau\bi v^\eta[\bi a^\eta-a^\xi].
\end{eqnarray}

We can obtain $E_{{\rm R}[\sigma(\eta)]\II,\II}^{\ \flat}$ in the 
same way as $E_{{\rm R}[\sigma(\eta)]\I,\I}^{\ \flat}$ and 
$E_{{\rm R}[\sigma(\eta)]\I,\II}^{\ \flat}$, although with a very 
laborious calculation (for an easier calculation, see Appendix \ref{new}).
Substituting ${T_{\II,\II}}^{\mu\nu}$ into Eq.~(\ref{ERindler}), and 
considering the limit  $\eta\rightarrow\infty$, The integration gives 
\begin{eqnarray}
\hspace{-0.8cm}\lim_{\eta\rightarrow\infty}
E^\flat_{{\rm R}[\sigma(\eta)]\II,\II}&=&
e^2d\tau\bi^2\left[
\frac{2}{3}v^\eta(a^\mu a_\mu)-\frac{2}{3}a^\eta(v^\mu a_\mu)
-2v^\eta(v^\mu a_\mu)^2
\right].
\label{mukakou}
\end{eqnarray}
We note here that this result is obtained without using any property 
peculiar to the 4-acceleration $a^\mu$ $[$as in Eq.~(\ref{I-II})$]$. 
By now using the property $a^\mu v_\mu=0$ of the 4-acceleration $a^\mu$, 
we obtain
\begin{eqnarray}
\lim_{\eta\rightarrow\infty}
E^\flat_{{\rm R}[\sigma(\eta)]\II,\II}&=&
\frac{2e^2}{3}\bi^2v^\eta(a^\mu a_\mu)d\tau.
\end{eqnarray}

Now we have obtained each contribution to 
$E^\flat_{{\rm R}[\sigma(\eta)]}$ due to the part that is 
independent of, linear in, or quadratic in the acceleration.
To obtain the energy $E_{{\rm R}[\sigma(\eta)]}$ with respect to the 
Killing field with an arbitrary norm, we simply multiply 
$E^\flat_{{\rm R}[\sigma(\eta)]}$ by $\nu$. By using 
Eqs. (\ref{uR}), (\ref{gR}) and 
$\nu\bi^2v^\eta=\nu(-\bar{\partial_\eta}\cdot v)=(-\bar{X}\cdot v)$, 
we can rewrite the results in covariant forms to obtain 
Eqs. (\ref{formI_I}) -- (\ref{formII_II}).

\section{Simple derivation of Eq.~(\protect\ref{formII_II})}
\label{new}

By using the property that the energy of the emitted part propagates 
along the future light cone without damping, one can obtain  
$E_{{\rm R}[\sigma(\eta)]\II,\II}$ in a simpler manner. Although, 
in Eq.~(\ref{yurui}), we have considered the energy only in the 
region where  $\Delta L$ intersects the plane        
$\tilde{\eta}=[{\rm constant}]$, we can obviously generalize 
this region for
the intersection of $\Delta L$ with arbitrary space-like surfaces. 
That is, for the regions $\sigma_1$ and $\sigma_2$ which are the 
intersections of $\Delta L$ with space-like surfaces $\Sigma_1$ and 
$\Sigma_2$, it follows that
\begin{eqnarray}
E_{{\rm R}[\sigma_1]\II,\II}&=&E_{{\rm R}[\sigma_2]\II,\II}.
\end{eqnarray}
By using this property, we can calculate 
$E_{{\rm R}[\sigma(\eta)]\II,\II}$ by selecting the region of 
integration in which $R$ is a constant. This can be done by cutting the 
region  $\Delta L$ with the simultaneous plane of the inertial frame 
in which the charge is instantaneously at rest at the radiation point 
$\bar{p}$.

Since the Rindler frame is symmetric with respect to rotation within the   
$x$-$y$ plane, we can put $v^y=0$ without loss of generality. Let us 
consider an inertial frame 
$\widetilde{\rm M}(\tilde{t},\tilde{x},\tilde{y},\tilde{z})$ 
that moves in the direction of the $z$ axis of the inertial frame, 
${\rm M}(t,x,y,z)$. Two inertial frames are related by the Lorentz 
transformation
\begin{eqnarray}
\left(
\begin{array}{c}
\tilde{t}\\
\tilde{z}
\end{array}
\right)
&=&
\left(
\begin{array}{cc}
\cosh\omega&\sinh\omega\\
\sinh\omega&\cosh\omega
\end{array}
\right)
\left(
\begin{array}{c}
t\\
z
\end{array}
\right).
\end{eqnarray}
Here we define the Rindler frame ${\rm R}(\eta,x,y,\xi)$ for the 
inetial frame
${\rm M}$ with  $t=\xi\sinh\eta$ and $z=\xi\cosh\eta$, and the 
Rindler frame
$\widetilde{\rm R}(\tilde{\eta},\tilde{x},\tilde{y},\tilde{\xi})$
for the inertial frame $\widetilde{\rm M}$ with 
$\tilde{t}=\tilde{\xi}\sinh\teta$ and 
$\tilde{z}=\tilde{\xi}\cosh\teta$.
Then we have a transformation between two Rindler frames 
\begin{eqnarray}
\tilde{\eta}&=&\eta+\omega\nonumber\\
\tilde{\xi}&=&\xi,
\label{etaomega}
\end{eqnarray}
which expresses the well-known fact that a Lorentz boost in the $z$ 
direction in the inertial frame corresponds to a time translation in 
the Rindler frame. We find that 
$\partial_{\tilde{\eta}}=\partial_{\eta}$ from Eq.~(\ref{etaomega}). 
Here we choose the frame $\widetilde{\rm M}$ so that $v^{\tilde{z}}=0$.
We also have $v^{\tilde{y}}=0$, because $v^y=0$. Therefore, only the
$\tilde{x}$ component of the velocity remains.  Now, by applying a Lorentz 
boost to the frame $\widetilde{\rm M}$ in the $\tilde{x}$ direction, we set 
the inertial frame ${\rm M}_0(T,X,Y,Z)$, in which the charge is 
instantaneously at rest at $\bar{p}$:
\begin{eqnarray}
T&=&v^{\tilde{t}}\tilde{t}-v^{\tilde{x}}\tilde{x},\ \ \ 
X=-v^{\tilde{x}}\tilde{t}+v^{\tilde{t}}\tilde{x},\ \ \ 
Y=\tilde{y},\ \ \ Z=\tilde{z},\\
v^T&=&1,\ \ \ v^X=v^Y=v^Z=0.
\end{eqnarray}

We perform the integral $E_{{\rm R}[\sigma]\II,\II}$ over the 
intersection
$\sigma(T)$ of the region $\Delta L$ with the plane $T=[{\rm constant}]$.
In the space $(X,Y,Z)$, $\sigma(T)$ is bounded by the sphere $S_0$ with 
radius $R^T=-v^\mu R_\mu=R$ centered at $(\bar{X},\bar{Y},\bar{Z})$, 
and the sphere $S'_0$ with radius $R-d\tau$ centered at 
$(\bar{X},\bar{Y},\bar{Z})$. Using the solid angle $d\Omega$  
centered at the point  
$(\bar{X},\bar{Y},\bar{Z})$, the integral is expressed as
\begin{eqnarray}
\hspace{-1cm}E_{{\rm R}[\sigma]\II,\II}&=&
-d\tau \int_{S_0} 
d\Omega R^2 n_\mu X_\nu {T_{\II,\II}}^{\mu\nu}\nonumber\\
\hspace{-1cm}&=&
-d\tau\int_{S_0} d\Omega \frac{e^2}{4\pi R^4}
[(a^\alpha R_\alpha)^2-R^2(a^\alpha a_\alpha)]
(n^\mu R_\mu)(X^\nu R_\nu).
\label{su}
\end{eqnarray}
From $X^\mu=\nu{\partial_\eta}^\mu$ and
\begin{eqnarray}
\partial_\eta&=&\partial_{\tilde{\eta}}
=\tilde{z}\partial_{\tilde{t}}+\tilde{t}\partial_{\tilde{z}}
\nonumber\\
&=&Z(
v^{\tilde{t}}\partial_T-v^{\tilde{x}}\partial_X
)+
(v^{\tilde{t}}T+v^{\tilde{x}}X)\partial_Z,
\end{eqnarray}
we obtain 
$\nu^{-1}(X^\mu R_\mu)=({\partial_\eta}^\mu R_\mu)=
-v^{\tilde{t}}\bar{Z}R-v^{\tilde{x}}\bar{Z}R^X
+(v^{\tilde{t}}\bar{T}+v^{\tilde{x}}\bar{X})R^Z$.  We now substitute 
this equation, together with the equations $(n^\mu R_\mu)=-R$ 
and $a^T=-v^\mu a_\mu=0$, into Eq.~(\ref{su}), and perform the 
angular integration by using the formulae
\begin{eqnarray}
\int_{S_0} d\Omega\frac{R^i}{R}&=&
\int_{S_0} d\Omega\frac{R^i R^j R^k}{R^3}=0,\ \ \ 
\int_{S_0} d\Omega\frac{R^i R^j}{R^2}=
\frac{4\pi}{3}\delta^{ij}.\nonumber\\
 & & \hspace{5cm}(i,\ j,\ k=X,\ Y,\ Z)
\end{eqnarray}
Then we obtain
\begin{eqnarray}
E_{{\rm R}[\sigma]\II,\II}&=&
\frac{2e^2}{3}\nu\bar{Z}v^{\tilde{t}}(a^\mu a_\mu)d\tau.
\label{last}
\end{eqnarray}
By noting 
$\nu\bar{Z}v^{\tilde{t}}=\nu\tilde{\xi}^2|_{\bar{p}} v^{\tilde{\eta}}
=\nu(-\bar{\partial}_{\tilde{\eta}}\cdot v)=(-X\cdot v)$, 
we obtain Eq.~(\ref{formII_II}).

In the derivation of Eq.~(\ref{formII_II}) performed in the previous 
and present appendices, 
we have not used any condition peculiar to the 4-acceleration $a^\mu$ 
other than $a^\mu v_\mu=0$.  
Therefore, even if we replace  $a^\mu$ with an arbitrary vector that 
is orthogonal to $v^\mu$ in Eq.  (\ref{IIandII}), the result of 
the integration in (\ref{ERindler}) has the same form as 
Eq.~(\ref{formII_II}). Then, we find that the radiation power of  
part $\tII$ is also obtained by simply replacing $a^\mu$ in 
Eq.~(\ref{formII_II}) with $\alpha^\mu$, 
with the result (\ref{tIIrad}).

\section{Integration of equation $\alpha^\mu=\we^\mu$}
\label{details}

In this apppendix, we solve the set of equations (\ref{eqkappa}), 
(\ref{damping}), 
(\ref{eqx}) and (\ref{eqy}).  Integration of 
Eq.~(\ref{eqkappa}) gives
\begin{eqnarray}
v^\eta-\xi^{-1}v^\xi&=&\kappa,
\label{const}
\end{eqnarray}
where $\kappa$ is a constant of integration.
From the conditions $v^\mu v_\mu=-1$ and $v^\eta>0$ and the restriction 
$\xi>0$, we find $\kappa>0$. We rewrite this equation as
\begin{eqnarray}
d[\eta-\ln\xi -\kappa\tau]&=&0,
\end{eqnarray}
which is integrated to give 
\begin{eqnarray}
\xi&=&\zeta e^{\eta-\kappa\tau},
\label{xixi}
\end{eqnarray}
where $\zeta>0$ is a constant of integration. Integration of 
Eq.~(\ref{damping}) gives
\begin{eqnarray}
(\xi v^\eta)^2-1&=&Ae^{-2\eta},
\end{eqnarray}
where $A$ is a constant of integration. By substituting Eq.~(\ref{xixi}) 
into this equation, we can eliminate the variable $\xi$:
\begin{eqnarray}
d[\sqrt{e^{2\eta}+A}-(\kappa\zeta)^{-1}e^{\kappa\tau}]&=&0
\end{eqnarray}
Then we obtain Eq.~(\ref{soleta}),
where $B$ is a constant of integration. By using  Eq.~(\ref{soleta}), 
we can eliminate  $e^\eta$ in 
Eq.~(\ref{xixi}). Then we obtain Eq.~(\ref{solxi}).

Integration of Eqs. (\ref{eqx}) and (\ref{eqy}) is trivial. 
Substitution of Eq.~(\ref{const}) into Eqs. (\ref{eqx}) and 
(\ref{eqy}) gives
\begin{eqnarray}
\frac{dv^x}{d\tau}&=&-\kappa v^x,\ \ \ \ \frac{dv^y}{d\tau}=-\kappa v^y.
\end{eqnarray}
Then we obtain Eqs. (\ref{solx}) and (\ref{soly}), 
where $\beta_x$, $x_f$, $\beta_y$ and $y_f$ are constants of integration.

From Eq.~(\ref{const}) and by differentiating the solutions  
(\ref{soleta}), (\ref{solx}) and (\ref{soly}) by $\tau$, 
we obtain the relation 
between the components of 4-velocity 
\begin{eqnarray}
\hspace{-1cm}(\xi v^\eta)^2-(v^x)^2-(v^y)^2-(v^\xi)^2
&=&
1-\kappa^2 e^{-2\kappa\tau}[B^2-A\zeta^2+{\beta_x}^2+{\beta_y}^2].
\end{eqnarray}
Therefore, we find that the condition $v^\mu v_\mu=-1$ is satisfied 
if and only if Eq.~(\ref{proper_time}) holds.

\section{Comparison with the work of Fugmann and Kretzschmar}
\label{comparison}

There is an alternative attempt to generalize the work of Rohrlich 
and Teitelboim in inertial frames to general accelerated frames, which 
was made by Fugmann and Kretzschmar in 1991.~\cite{FK91} \  In this 
appendix, we translate their result into our notation, and compare 
this with our result.
 
According to their result, the radiation fields are given by a 
contribution to 
Li\'enard-Wiechert fields due to the part linear in 
${\bf \alpha_q}-{\bf b_\perp}$ 
$[$see Eq.~(6.2) in Ref.~\cite{FK91}$]$. Roughly 
speaking, ${\bf \alpha_q}$ represents the acceleration of the 
charge, and ${\bf b_\perp}$ expresses the acceleration and rotation 
of the frame.  

In the case of the Rindler frame, ${\bf b_\perp}$ is expressed in our 
notation in the form
\begin{eqnarray}
{b_\perp}^\mu&=&
[\delta^\mu_{\ \nu}-\hat{\eta}^\mu\hat{\eta}_\nu]g^\nu\nonumber\\
&=&
g^\mu-\frac{(g\cdot R)}{(u\cdot R)}u^\mu
-\frac{(g\cdot R)}{(u\cdot R)^2}R^\mu
\equiv
\beta^\mu-\frac{(g\cdot R)}{(u\cdot R)^2}R^\mu,
\end{eqnarray}
where $\hat{\eta}^\mu=-(u\cdot R)^{-1}R^\mu-u^\mu$, and this quantity 
is expressed as ${\bf n}$ in their paper. Their definition of the
radiation fields (given by Eq.~(6.9) in their paper) is translated 
into the following expression:
\begin{eqnarray}
{F_{_{\rm FK}}}^{\mu\nu}&=&
\frac{e}{R^3}(a^\alpha-{b_\perp}^\alpha)R_\alpha[v^\mu R^\nu-R^\mu v^\nu]
+\frac{e}{R^2}[(a^\mu-{b_\perp}^\mu)R^\nu-R^\mu(a^\nu-{b_\perp}^\nu)]
\nonumber\\
&=&\frac{e}{R^3}(a^\alpha-\beta^\alpha)R_\alpha[v^\mu R^\nu-R^\mu v^\nu]
+\frac{e}{R^2}[(a^\mu-\beta^\mu)R^\nu-R^\mu(a^\nu-\beta^\nu)],\nonumber\\
\label{FK1}
\end{eqnarray}
or
\begin{eqnarray}
{F_{_{\rm FK}}}^{\mu\nu}
&=&\frac{e}{R^3}h^\lambda_{\ \rho}(a^\rho-\beta^\rho)
R_\lambda[v^\mu R^\nu-R^\mu v^\nu]\nonumber\\
& &+\frac{e}{R^2}[h^\mu_{\ \rho}(a^\rho-\beta^\rho)R^\nu
-R^\mu h^\nu_{\ \rho}(a^\rho-\beta^\rho)].
\label{FK2}
\end{eqnarray}
When the charge is instantaneously at rest in the Rindler frame 
($v^\mu=u^\mu$), $h^\lambda_{\ \rho}(a^\rho-\beta^\rho)$ reduces to 
$a^\mu-g^\mu$. Therefore, in this special case, their result is 
identical to our result. Although, in more general situations, 
these results are not equivalent.

Although Eqs.~(\ref{FK1}) and (\ref{FK2}) resemble 
Eq.~(\ref{lambda}) in appearance, there is a crucial defference 
between these. Because $a^\mu-\beta^\mu$ depends on $R^\mu$, this 
term cannot be considered to be the vector at the point  $\bar{p}$, 
in contrast to the vector $\lambda^\mu$ in Eq.~(\ref{lambda}). 
Therefore, we cannot apply the result of \S~\ref{Kgeneral} to the 
present discussion. By using the formulae given in Eqs. (\ref{one}), 
(\ref{two}) and (\ref{tau}), and the property $\beta^\mu R_\mu=0$, 
we obtain the equation satisfied by ${F_{_{\rm FK}}}^{\mu\nu}$ off 
the world line of the charge,
\begin{eqnarray}
\nabla_{[\lambda}F_{{_{\rm FK}}\mu\nu]}
&=&\frac{2e}{R^3}\varepsilon_{[\lambda}R_\mu\beta_{\nu]},
\label{crucial}\\
\nabla_\mu {F_{_{\rm FK}}}^{\mu\nu}&=&
-\frac{eR^\nu}{R^4}[2h^\mu_{\ \alpha}(a^\alpha-\beta^\alpha)R_\mu
-R(\varepsilon\cdot\beta)]\label{propR},
\end{eqnarray}
where $\varepsilon^\mu=v^\mu+(u\cdot R)^{-1}Ru^\mu$ and 
$\varepsilon^\mu R_\mu=0$.

In general situations, the right-hand side of Eq.~(\ref{crucial}) 
does not vanish, in contrast to Eq.~(\ref{four_potential}) in 
\S~\ref{Kgeneral}. This fact implies that there are no four-potentials 
which express the field ${F_{_{\rm FK}}}^{\mu\nu}$. As long as we are 
concerned with Eqs. (\ref{crucial}) and (\ref{propR}), our identification 
of the radiation fields ${F_{\ \tII}}^{\mu\nu}$ seems to be more natural 
than their identification ${F_{_{\rm FK}}}^{\mu\nu}$, because our 
equations (\ref{four_potential}) and (\ref{Rnatural}) are considered to  
be a proper generalization of equations given by Teitelboim, with the 
replacement of $a^\mu$ and $\partial_t$ with $\alpha^\mu$ and $X^\mu$.
However, it might be possible to interpret that the generalization of 
the concept of radiation given by Fugmann and Kretzschmar is 
conceptually different from ours, because their identification is 
based on the use of advanced optical cooordinates (not
nesessarily stationary frames), and advanced optical coordinates 
might possess other symmetries that do not appear in our formulation.  
 
In any case, we can show that their identification of the radiation also 
satisfies conditions 1 and 2 introduced in \S~\ref{Kgeneral}, despite the
fact that the field ${F_{_{\rm FK}}}^{\mu\nu}$ is inequivalent to 
the field $f^{\mu\nu}$. Let us consider this point in the following.

First of all, we can easily confirm contition 1 by seeing that
${F_{_{\rm FK}}}^{\mu\nu}$ is equivalent to
${F_{\ \tII}}^{\mu\nu}$ when the charge is at rest in 
the Rindler frame ($v^\mu=u^\mu$).
We now consider condition 2. From the relation 
$F^{\mu\nu}=-F^{\nu\mu}$, we obtain
\begin{eqnarray}
4\pi \partial_\mu T^{\mu\nu}&=&
(\partial_\mu F^{\mu\alpha})F_\alpha^{\ \nu}
+\frac{3}{2}\partial^{[\nu}F^{\alpha\beta]}
F_{\alpha\beta}.
\end{eqnarray}
From ${F_{_{\rm FK}}}^{\mu\nu}R_\nu=0$ and Eq.~(\ref{propR}), 
we find  
$(\partial_\mu {F_{_{\rm FK}}}^{\mu\alpha})
{F_{_{\rm FK}}}_\alpha^{\ \nu}=0$.
Here, $\varepsilon^\mu$, $R^\mu$ and $\beta^\mu$ are orthogonal 
to $R^\mu$. Then we have 
$\partial^{[\nu}{F_{_{\rm FK}}}^{\alpha\gamma]}
{F_{_{\rm FK}}}_{\alpha\gamma}
=(2e/R^3)\varepsilon^{[\nu}R^\alpha\beta^{\gamma]}
{F_{_{\rm FK}}}_{\alpha\gamma}
=0$. From these result, we obtain the conservation law
\begin{eqnarray}
\partial_\mu {T_{_{\rm FK}}}^{\mu\nu}&=&0.
\end{eqnarray}
From this and ${T_{_{\rm FK}}}^{\mu\nu}\propto R^\mu R^\nu$,
we find that ${T_{_{\rm FK}}}^{\mu\nu}$ satisfies Eq.~(\ref{strong}),
which is the mathematical expression of condition 2.

\end{document}